\documentclass[journal]{IEEEtran}

\usepackage{calc,amsfonts,amssymb,amsmath,bm,xurl,color,graphicx,cite,epstopdf,nicefrac,bbold, amsthm}
\usepackage{psfrag,subfigure,float}
\usepackage[algoruled,linesnumbered]{algorithm2e}
\usepackage{multirow}

\usepackage{algorithmic}
\usepackage[normalem]{ulem}
\usepackage{stmaryrd}
\usepackage{xcolor}
\usepackage{psfrag,framed,mdframed}
\usepackage{comment}
\usepackage{mdframed}
\usepackage{cleveref}
\usepackage[utf8]{inputenc}
\usepackage{tikz}

\usepackage{mathtools}

\definecolor{green}{RGB}{50,150,50}
\definecolor{orange}{RGB}{255,107,0}
\definecolor{purple}{RGB}{255, 0, 255}

\makeatletter
\def\endthebibliography{%
	\def\@noitemerr{\@latex@warning{Empty `thebibliography' environment}}%
	\endlist
}
\makeatother

\makeatletter
\newcommand{\nosemic}{\renewcommand{\@endalgocfline}{\relax}}% Drop semi-colon ;
\newcommand{\dosemic}{\renewcommand{\@endalgocfline}{\algocf@endline}}% Reinstate semi-colon ;
\let\oldnl\nl% Store \nl in \oldnl
\newcommand{\nonl}{\renewcommand{\nl}{\let\nl\oldnl}}% Remove line number for one line
\makeatother

\newlength\myindent
\usepackage{steinmetz}
\renewcommand{\a}{\boldsymbol{a}}

\newcommand{\e}{\boldsymbol{e}}
\newcommand{\f}{\boldsymbol{f}}
\newcommand{\g}{\boldsymbol{g}}
\newcommand{\h}{\boldsymbol{h}}
\newcommand{\m}{\boldsymbol{m}}
\newcommand{\n}{\boldsymbol{n}}

\newcommand{\q}{\boldsymbol{q}}

\newcommand{\w}{\boldsymbol{w}}
\newcommand{\y}{\boldsymbol{y}}
\newcommand{\x}{\boldsymbol{x}}
\newcommand{\z}{\boldsymbol{z}}

\newcommand{\btheta}{\boldsymbol{\theta}}

\newcommand{\bpi}{\boldsymbol{\pi}}
\newcommand{\zero}{\boldsymbol{0}}
\newcommand{\one}{\boldsymbol{1}}
\newcommand{\I}{\boldsymbol{I}}

\newcommand{\W}{\boldsymbol{W}}

\newcommand{\Q}{\boldsymbol{Q}}
\newcommand{\X}{\boldsymbol{X}}

\newcommand{\C}{\boldsymbol{C}}

\renewcommand{\H}{\boldsymbol{H}}

\newcommand{\Z}{\boldsymbol{Z}}

\newcommand{\T}{{\!\top\!}}
\newcommand{\cA}{\mathcal{A}}
\newcommand{\cB}{\mathcal{B}}
\newcommand{\cC}{\mathcal{C}}
\newcommand{\cD}{\mathcal{D}}
\newcommand{\cE}{\mathcal{E}}
\newcommand{\cF}{\mathcal{F}}
\newcommand{\cG}{\mathcal{G}}
\newcommand{\cH}{\mathcal{H}}

\newcommand{\cL}{\mathcal{L}}

\newcommand{\cN}{\mathcal{N}}
\newcommand{\cO}{\mathcal{O}}

\newcommand{\cR}{\mathcal{R}}
\newcommand{\cS}{\mathcal{S}}
\newcommand{\cT}{\mathcal{T}}
\newcommand{\cU}{\mathcal{U}}

\newcommand{\cX}{\mathcal{X}}

\newcommand{\bbR}{\mathbb{R}}
\newcommand{\bbC}{\mathbb{C}}
\newcommand{\bbE}{\mathbb{E}}

\DeclareMathOperator*{\minimize}{\textrm{minimize}}

\newtheorem{theorem}{Theorem}
\newtheorem{assumption}{Assumption}

\newtheorem{lemma}{Lemma}

\newtheorem{definition}{Definition}
\newtheorem{remark}{Remark}

\usepackage{xcolor}
\usepackage{psfrag,framed}
\usepackage{lipsum}
\usepackage{chapterbib}
\PassOptionsToPackage{normalem}{ulem}
\usepackage{ulem}
\definecolor{shadecolor}{RGB}{220,220,220}
\begin{document}
	
	% \begin{bibunit}[IEEEtran]
		\title{Optimal Solutions for Joint Beamforming and Antenna Selection: From Branch and Bound to {Graph Neural Imitation Learning}}
		
		\author{Sagar Shrestha, Xiao Fu, and Mingyi Hong
			\thanks{S. Shrestha and X. Fu are with the School of EECS, Oregon State University. M. Hong is with the ECE Department, University of Minnesota. ({\it Corresponding Author: Xiao Fu})
				
				The work of S. Shrestha and X. Fu is supported by National Science Foundation (NSF) under Project CNS-2003082. Their work is also supported by a gift from Intel through the MLWiNS program. The work of M. Hong is supported by NSF CNS-2003033. His work is also supported by a gift from Intel through the MLWiNS program. 
		}}
		
		\maketitle
		\begin{abstract}
			This work revisits the joint beamforming (BF) and antenna selection (AS) problem, as well as its robust beamforming (RBF) version under imperfect channel state information (CSI). 
			Such problems arise due to various reasons, e.g., the costly nature of the radio frequency (RF) chains and energy/resource-saving considerations.
			The joint (R)BF\&AS problem is a mixed integer and nonlinear program, and thus finding {\it optimal solutions} is often costly, if not outright impossible. The vast majority of the prior works tackled these problems using techniques such as continuous approximations, greedy methods, and supervised machine learning---yet these approaches do not ensure optimality or even feasibility of the solutions. 
			The main contribution of this work is threefold. First, an effective {\it branch and bound} (B\&B) framework for solving the problems of interest is proposed.
			Leveraging existing BF and RBF solvers, it is shown that the B\&B framework guarantees global optimality of the considered problems. 
			Second, to expedite the potentially costly B\&B algorithm, a machine learning (ML)-based scheme is proposed to help skip intermediate states of the B\&B search tree.
			The learning model features a {\it graph neural network} (GNN)-based design that is resilient to a commonly encountered challenge in wireless communications, namely, the change of problem size (e.g., the number of users) across the training and test stages.
			Third, comprehensive performance characterizations are presented, showing that the GNN-based method retains the global optimality of B\&B with provably reduced complexity, under reasonable conditions. 
			Numerical simulations also show that the ML-based acceleration can often achieve an order-of-magnitude speedup relative to B\&B. 
		\end{abstract}
		
		\IEEEpeerreviewmaketitle

		\section{Introduction}
		Beamforming lies at the heart of transmit signal design of multiple antenna systems.
		In the past decade, a plethora of beamforming algorithms have been proposed under various scenarios; see, e.g., 
		\cite{golbon2016beamforming,sidiropoulos2006transmit,gershman2010convex,wang2014outage,mehanna2013joint,shi2014group,ibrahim2020fast}.
		Among the most challenging scenarios is the joint beamforming and antenna selection (BF\&AS) problem (see, e.g., \cite{mehanna2013joint,ibrahim2020fast,shi2014group}), which often arises due to various reasons---such as the costly nature of radio frequency (RF) chains \cite{marinello2020antenna, sanayei2004antenna, gao2017massive, mehanna2013joint}, energy consumption considerations \cite{jiang2012antenna, arora2021analog}, problem size reduction \cite{molisch2005capacity}, overhead minimization \cite{liu2014joint}, and algorithm accommodations \cite{sadek2007active}.

		Jointly designing the beamformers and selecting antennas is a {\it mixed integer and nonlinear program}, which is known to be NP-hard \cite{konar2018simple, civril2009selecting}. A large portion of the literature tackles this problem using continuous programming-based approximations.
		For example, \cite{mehanna2013joint, ahmed2020antenna,shi2014group,ibrahim2020fast} used convex and nonconvex group sparsity-promoting regularization to encourage turning off antenna elements. 
		However, the continuous approximations are often NP-hard problems as well (especially when the sparsity promotion is done via nonconvex quasi-norms as in \cite{mehanna2013joint}), and thus it is unclear if they can solve the problem of interest optimally. 
		In addition, works using greedy methods to assist antenna selection also exist (see, e.g., \cite{chen2007efficient, mendoncca2019antenna, ding2010mmse, mahdi2021quantization,konar2018simple}). But the optimality of joint (R)BF\&AS is still not addressed in these works.
		
		In recent years, machine learning (ML) approaches are employed to handle the joint BF and AS problem. In \cite{ibrahim2018learning}, a supervised learning approach was proposed. The basic idea is to use a continuous optimization algorithm to produce training pairs (i.e., channel matrices and sparse beamformers), and then learn a neural network-based regression function using such pairs. Similar ideas were used in \cite{vu2021machine, elbir2019joint} with various settings. This type of approach in essence mimics the training pair-generating algorithms at best, and thus the optimality of their solutions is again not guaranteed.
		
		\noindent{\bf Contributions.}
		In this work, we revisit the joint BF and AS and its extension under imperfect channel state information (CSI), namely, the joint robust beamforming (RBF) and AS problem.
		We are interested in the unicast BF and RBF formulations in {\cite{bengtsson2001optimum} } and \cite{zheng2008robust}, respectively. The goal is to satisfy the users' quality-of-service (QoS) constraints while minimizing the power consumption, with only a subset of the antenna elements activated. Our detailed contributions are as follows:
		
		\noindent
		$\bullet$ {\bf Optimal Joint (R)BF\&AS via Branch and Bound.} Our first contribution lies in an optimal computational framework to attain the {\it global optimal solutions} to the joint (R)BF\&AS problems. To this end, we propose a {\it Branch and Bound} (B\&B) \cite{clausen1999branch, land1960automatic} framework that is tailored for the problems of interest. Our design leverages problem structures of unicast BF and RBF, which allows 
		for branching only on a subset of the optimization variables---thereby having reduced complexity and being effective. Unlike continuous optimization-based approximations in \cite{mehanna2013joint, ahmed2020antenna,shi2014group,ibrahim2020fast} whose solutions are often sub-optimal or infeasible, the proposed B\&B is guaranteed to return an optimal solution.
		
		\noindent
		$\bullet$ {\bf An ML-based Acceleration Scheme.} B\&B is known for its relatively weak scalability. To improve efficiency, an idea from the ML community (see, e.g., \cite{he2014learning, nair2020solving}) is to learn a binary classifier offline using multiple problem instances. The classifier determines whether or not any encountered intermediate state of the B\&B algorithm could be ``skipped'', as skipping these states saves computational resources and expedites the B\&B process. 
		Generic ML learning functions (e.g., support vector machines (SVM)) used in existing works like \cite{he2014learning, lee2019learning} do not reflect the problem structure in wireless communications.
		In this work, we propose a {\it graph neural network} (GNN) \cite{scarselli2008graph} based learning function designed to exploit the physics of the (R)BF problem---which offers an enhanced classification accuracy. More importantly, the GNN is agnostic to the change of scenarios (e.g., problem size) during training and testing. This feature is designed to meet the need of wireless communication systems, as the number of users served by a base station could change quickly in practice.

		\noindent
		$\bullet$ {\bf Theoretical Understanding.} We present comprehensive performance characterizations for the proposed approaches. In particular, we show that the ML-based acceleration retains the global optimality of the B\&B procedure with high probability, under reasonable conditions. 
		ML-based B\&B acceleration has limited theoretical studies, and the results were developed under often overly ideal settings (e.g., convex classifier) \cite{he2014learning, ross2011reduction}. There is a lack of understanding of the impacts of key factors such as nonconvexity, limited training samples, and the employed ML model's structure.
		Our analysis takes {important aspects into consideration, }such as the nonconvexity of the GNN learning process, the GNN's structure and complexity, the GNN's function approximation error, and the amount of available samples. As a consequence, the analysis offers insights to reveal key trade-offs in practice.

		\smallskip

		A shortened version was submitted to ICASSP 2023 \cite{icassp2023submission}. The conference version included the B\&B design and ML acceleration for the perfect CSI case. The journal version additionally presents the imperfect CSI case, the detailed analysis of the B\&B algorithm, the performance characterizations for the ML acceleration, and more extensive simulations.
		
		\noindent{\bf Related Works.} B\&B was proposed for beamforming problems in \cite{lu2017efficient, lu2020enhanced}, and antenna selection problems in \cite{ouyang2019optimal, li2014globally, gao2015bidirectional}.
		Particularly, the work in \cite{lu2017efficient} considered a single group multicast beamforming problem, the work in \cite{li2014globally} considered a joint power allocation and antenna selection problem, the work in \cite{ouyang2019optimal} considered antenna selection-assisted rate maximization in wiretap channels, and \cite{gao2015bidirectional, gao2017massive} considered receive antenna selection for sum rate maximization. 
		{However these are different from the QoS-constrained downlink transmit beamfroming formulation}
		considered in our work, which requires new B\&B designs.
		ML-based B\&B acceleration so far has been mostly used for \textit{mixed integer and linear programs} (MILPs) in the ML community, e.g., \cite{he2014learning, gasse2019exact}, where the B\&B design is standard. Such methods have also been adopted in wireless communications in \cite{lee2019learning, shen2019lorm} where resource allocation tasks are framed as \textit{mixed integer and nonlinear programs} (MINPs). However, the joint (R)BF\&AS problem has not been considered. In addition, comprehensive theoretical understanding to such ML-acceleration procedures has been elusive.

		\smallskip
		
		\noindent
		\textbf{Notation:} $x$, $\x$ and $\X$ denote a scalar, a vector, and a matrix, respectively. $\x_n$ denotes the $n$th column of $\X$. We use the matlab notation $\X(n,:)$ to denote the $n$th row of $\X$. $[N]$ denotes the set $\{1, 2, \dots, N\}$. $\|\x\|_2$, $\|\x\|_{\infty}$, $\|\X\|_2$, $\|\X\|_F$, $\|\X\|_{\rm row-0}$ denote the vector $\ell_2$ norm, vector $\ell_\infty$ norm, matrix spectral norm, matrix Frobenius norm, and the number of non-zero rows in the matrix, respectively. ${\rm Tr}(\X)$, $\X^H$, and $\X^\T$ denote the trace, hermitian, and transpose of $\X$. $|\cX|$ denotes the cardinality of the set $\cX$. $\bbE [\cdot]$ denotes the expectation operator. $\X \succeq \bm 0$ denotes that $\X$ is positive semi-definite matrix. $\X({\cS},:)$ with $\cS \subseteq [N]$ denotes the submatrix of $\X \in \bbC^{N \times M}$ containing only the rows of $\X$ contained in the set $\cS$. $\X_{-n}$ denotes the submatrix of $\X$ with the $n$th column removed. $\f(\cdot)$ is $C$-Lipschitz continuous iff $\|\f(\x) - \f(\y)\|_2 \leq C\|\x - \y\|_2$.
		
		\section{Background}
		Consider a classic {single-cell} downlink communication scenario where the \textit{base station} (BS) has $N$ antennas {\cite{bengtsson2001optimum,ma2017unraveling}}. The BS serves $M$ single antenna users. 
		Denote
		$\w_m \in \bbC^N$ as the beamforming vector for serving user $m$. In this work, our interest lies in a scenario where the BS only activates $L$ anntennas. 
		Hence, we aim to select a subset ${\cal A}\subseteq \{1,\ldots,N\}$ such that $|{\cal A}|\leq L$ and $\w_m(i)\neq 0$ only when $i\in {\cal A}$.
		The message signal for user $m$ is represented by $s_m(t)$. Given the channel $\h_m \in \bbC^{N}$ between the BS and user $m$, the signal received by user $m$ can be expressed as follows:
		$$ y_m(t) = \h_m^H \w_m s_m(t) + \sum_{\ell \not = m} \h_m^H \w_\ell s_\ell(t) + \n_m,$$
		where $\n_m$ is zero-mean circular symmetric white Gaussian noise with variance $\sigma_m^2$. 
		Assume w.l.o.g. that $\{ s_m(t) \}_{m=1}^M$  are mutually uncorrelated and temporally white with zero-mean and unit-variance. Then, the total transmission power is given by
		$ \sum_{m=1}^M \|\w_m\|_2^2 := \| \W \|_{\rm F}^2,$
		where $\W = [\w_1, \dots, \w_M]$. The \textit{signal to interference and noise ratio} (SINR) at the $m$th receiver is expressed as:
		\begin{align}\label{eq:sinr}
			{\rm SINR}_m = \frac{|\w_m^H \h_m|_2^2 }{\sum_{\ell \not = m} | \w_\ell^H \h_m|_2^2 + \sigma_m^2}.
		\end{align}
		
		\subsection{Unicast Beamforming and SOCP}
		One of the most popular formulations for beamforming
		is the so-called QoS formulation \cite{rashid1998transmit, visotsky1999optimum, karipidis2008quality} that tries to maintain a pre-specified SINR level for all users. When $\h_m$ is known, the unicast BF problem can be formulated as follows:
		\begin{subequations}\label{eq:beamforming}
			\begin{align}
				\minimize_{\W} ~ &\|\W\|_F^2  \\
				\text{subject to } ~ & \frac{|\w_m^H\h_m|^2}{\sum_{\ell \not = m} |\w_\ell^H \h_m|^2 + \sigma_m^2 } \geq \gamma_m, \quad m \in [M]. \label{eq:sinr_bf}
			\end{align}
		\end{subequations}
		Problem~\eqref{eq:beamforming} is called ``unicast'' BF because every user receives its own message.
		Problem~\eqref{eq:beamforming} appears to be
		nonconvex, but it can be recast as a {\it second-order cone program} (SOCP):
		\begin{lemma}[\cite{bengtsson2001optimum}]\label{lemma:socp}
			Eq.~\eqref{eq:sinr_bf} can be equivalently written as a second-order cone constraint:
			\begin{align}\label{eq:bf_convex_reformulation}
				\frac{1}{\sqrt{\gamma_m \sigma_m^2}} {\rm Re}(\w_m^H\h_m) \geq \sqrt{\sum_{\ell \not = m} |\w_\ell^H \h_m|^2 + 1},
			\end{align}
			for all $m\in[M]$.
			Therefore, any {algorithm for solving SOCP} can be used to solve \eqref{eq:beamforming} optimally. 
		\end{lemma} %{\orange [Rev 3 other Comment 2]} {\red can keep our original ref here as well}{\orange That reference was a tutorial paper from the same author several years later. So I guess this is sufficient?}

		\subsection{Robust Beamforming and SDR}
		When the BS only has imperfect CSI, the following worst-case RBF formulation is often considered 
		\cite{zheng2008robust, kim2008robust, song2012robust, chang2011worst, ma2017unraveling}: 
		\begin{subequations}
			\label{eq:robust_beamforming}
			\begin{align}
				\minimize_{\W} ~ &\|\W\|_F^2  \\
				\text{subject to } ~ & \min_{\overline{\h}_m \in \cU_m} \frac{|\w_m^H \overline{\h}_m|^2}{\sum_{\ell \not = m} |\w_\ell^H \overline{\h}_m|^2 + \sigma_m^2 } \geq \gamma_m, \nonumber \\
				&\quad \quad  \quad \quad \quad \quad \quad  \quad \quad \quad \forall m \in [M], \label{eq:sinr_rbf} 
				% & \| \widetilde{\w} \|_0 \leq L. ,
			\end{align} 
		\end{subequations}
		where $\cU_m := \{ \h_m + \e_m ~|~  \|\e_m\|_2 \leq  \varepsilon_m \}$, $\h_m$ is the approximate channel vector available at the BS, and $\varepsilon_m$ is the bound on the approximation error. Problem~\eqref{eq:robust_beamforming} cannot be directly converted to a convex program as in the perfect CSI case (cf. Lemma~\ref{lemma:socp}). However, Problem~\eqref{eq:robust_beamforming} can be tackled by a convex relaxation technique, namely, {\it semidefnite relaxation} (SDR) \cite{luo2010semi}. Let $\W_m := \w_m \w_m^H$. Then the SDR of \eqref{eq:robust_beamforming} is given by
		\begin{subequations}\label{eq:robust_sdr}
			\begin{align}
				\minimize_{\{\W_m \in \bbC^{N \times N}\}_{\m=1}^M } ~ & \sum_{i=1}^M {\rm Tr}(\W_m) \\
				\text{subject to } ~ & \min_{\overline{\h}_m \in \cU_m} \frac{\overline{\h}_m^H \W_m \overline{\h}_m}{\sum_{j \not= m} \overline{\h}_m^H \W_j \overline{\h}_m + \sigma_m^2} \geq \gamma_m, \label{eq:sdr_sinr}  \\
				& \W_m \succeq \bm 0, \quad \forall m \in [M].\nonumber
			\end{align} 
		\end{subequations}
		Note that \eqref{eq:robust_sdr} and \eqref{eq:robust_beamforming} are equivalent if the constraint $\W_m = \w_m \w_m^H$ (or, ${\rm rank}(\W_m)=1$) has not been relaxed. {Problem} ~\eqref{eq:robust_sdr} can be further re-expressed as a standard semidefinite program (SDP) using the $S$-Lemma; see details in \cite{ma2017unraveling}.
		Interestingly, this relaxation procedure turns out to be tight under reasonable conditions:
		\begin{lemma}[{\cite[Theorem 1]{ma2017unraveling}}]\label{lemma:optimal_sdr}
			Suppose that Problem \eqref{eq:robust_beamforming} is feasible. 
			Let ${\bm \Pi}_{m} := \I - {\H}_{-m} ({\H}_{-m}^H {\H}_{-m})^{-1} {\H}_{-m}^H $ be the orthogonal complement projector of ${\H}_{-m}$. If 
			\begin{align}\label{eq:cond_rbf}
				\frac{\|\bm \Pi_{m} {\h}_m\|^2_2}{\varepsilon^2_m} > 1 + M + (M - \frac{1}{M}) \gamma_m, \forall m ,    
			\end{align}
			then the optimal solution of \eqref{eq:robust_beamforming} can be obtained using SDR.
		\end{lemma}
		The condition in \eqref{eq:cond_rbf} means that if the downlink channels associated with different users are sufficiently different, then the SDR is tight.

		\subsection{Joint (R)BF\&AS: Existing Approaches}\label{sec:existing_approaches}
		
		The joint (R)BF\&AS problem frequently arises in practice for many reasons. 
		For example, due to the the costly and power-hungry nature of RF chains, in some antenna arrays, the number of RF chains may be fewer than that of the antenna elements \cite{sanayei2004antenna, marinello2020antenna, gao2017massive, mehanna2013joint}. 
		Furthermore, AS is also used for energy-efficiency considerations \cite{jiang2012antenna}, problem size reduction, overhead control, and algorithm design accommodations---see, e.g., \cite{sadek2007active, molisch2004mimo, sanayei2004antenna,molisch2005capacity,liu2014joint} and the discussions therein.
		The problem considered in this work is as follows:
		\begin{mdframed}
			\begin{subequations}\label{eq:metaproblem}
				\begin{align}
					\minimize_W&~\|\bm W\|_F^2  \\
					{\rm subject~to}&~{\cal C}(\bm w_m,\bm h_m, \varepsilon_m, \sigma_m) \geq \gamma_m,  \label{eq:sinr_constraint} \\
					&~\|\bm W\|_{{\rm row}\text{-}0}\leq L.  \label{eq:cardinality}
				\end{align}
			\end{subequations}
		\end{mdframed}
		where the row-0 function $\|\cdot\|_{{\rm row}\text{-}0}$ counts the number of nonzero rows in $\W$ and
		\begin{align*}
			&{\cal C}(\bm w_m,\bm h_m, \varepsilon_m, \sigma_m) \\
			& :=  \begin{cases} \frac{|\w_m^H\h_m|^2}{\sum_{\ell \not = m} |\w_\ell^H \h_m|^2 + \sigma_m^2 }, \quad \text{if BF is considered,}\\
				\min_{\overline{\h} \in \cU_m}  \frac{|\w_m^H \overline{\h}_m|^2}{\sum_{\ell \not = m} |\w_\ell^H \overline{\h}_m|^2 + \sigma_m^2 }, \quad \text{if RBF is considered}.
			\end{cases}
		\end{align*}
		Problem~\eqref{eq:metaproblem} is a non-convex combinatorial problem, and it is NP-hard \cite{civril2009selecting}. Some representative approaches for tackling joint (R)BF\&AS problems are as follows:

		\subsubsection{Continuous Approximations}
		In the literature, Problem~\eqref{eq:metaproblem} and other joint (R)BF\&AS formulations are often handled by continuous approximation.
		For example, a representative continuous approximation technique was used in \cite{mehanna2013joint} for handling a multicast version of \eqref{eq:metaproblem}. Using the idea from \cite{mehanna2013joint}, one can recast the unicast problem in \eqref{eq:metaproblem} as a regularized formulation as follows:
		\begin{align}\label{eq:beamforming_as}
			\minimize_{\W} ~ &\|\W\|_F^2  + \lambda\|\W\|_{{\rm row}\text{-}0} \\
			\text{subject to } ~ & \cC(\w_m, \h_m, \varepsilon_m, \sigma_m) \geq \gamma_m, \quad m \in [M].\nonumber
		\end{align}
		The idea in \cite{mehanna2013joint} is to approximate the row-0 function using a group sparsity-inducing norm, namely, the $\ell_{\infty,1}$ norm, i.e., $\|\W\|_{{\rm row}\text{-}0}  \approx  \sum_{n=1}^N \|\W(n,:)\|_\infty$ and its nonconvex counterpart $\|\W\|_{{\rm row}\text{-}0}  \approx  \sum_{n=1}^N  \log\left(\|\W(n,:)\|_\infty+\varepsilon \right)$ \cite{candes2008enhancing}.
		Similar ideas were used in \cite{ahmed2020antenna}. 
		Such continuous approximations allow the use of standard nonlinear program techniques to tackle \eqref{eq:beamforming_as}.
		However, as mentioned, these methods do not provide any optimality guarantees. In addition, the feasiblity of $\W$ is often not met by the approximate solutions.

		\subsubsection{Greedy Methods} A number of greedy approaches also exist for tackling various formulations of the joint (R)BF\&AS problem; see, e.g., \cite{chen2007efficient, mendoncca2019antenna, ding2010mmse, mahdi2021quantization,konar2018simple}. The major idea is to activate or shut down an antenna in every iteration using a certain criterion that is often defined by the optimization problem's objective function---see an example in Sec.~\ref{sec:baselines}. Notably, such greedy algorithms are not necessarily computationally light, as will be seen in our simulations.
		
		\subsubsection{Supervised Learning}
		More recently, a number of learning-based approaches are proposed to tackle the joint (R)BF\&AS problem; see, e.g., \cite{ibrahim2018learning, joung2016machine, chen2019intelligent}. In \cite{ibrahim2018learning}, a multicast version of \eqref{eq:metaproblem} was considered. The idea is to use an existing joint multicast BF\&AS algorithm (e.g., the algorithm from \cite{mehanna2013joint}) to generate ``training pairs'' $ \{  \bm H_t,  \widehat{\bm W}_t \}_{t=1}^T $ by simulating a large number of problem instances,
		where $t$ is the instance index, $\widehat{\bm W}_t$ is a (row-sparse) solution produced by the training pair-generating algorithm, and $\H_t$ is the channel matrix of instance $t$.
		Note that the training pairs can take other forms, e.g., $\{\H_t,\widehat{\z}_t\}$ where $\widehat{\z}_t\in\mathbb{R}^M$ is a binary vector found by the training pair-producing algorithm, indicating which antenna is activated \cite{ibrahim2018learning, vu2021machine}.
		Then, a deep neural network (DNN) $\bm f_{\bm \theta}(\cdot)$ is trained via
		\begin{align}
			\widehat{\bm \theta}\leftarrow\arg\min_{\bm \theta}~\frac{1}{T}\sum_{t=1}^T \ell( \widehat{\bm W}_t, \bm f_{\bm \theta}(\bm H_t)   ),
		\end{align}
		where $\bm \theta$ represents the parameters of the DNN and $\ell(x,y)$ measures the divergence between $x$ and $y$.
		When a new $\H$ is seen in the test stage, one can use the learned DNN to predict the solution, i.e., $\widehat{\W} = \bm f_{\widehat{\bm \theta}}(\H)$. 
		This ``supervised learning'' idea is similar to a line of work in deep learning-based wireless system design; see, e.g., \cite{lee2018deep,sun2018learning}. Notably, it cannot exceed the performance of the algorithm that produces the training pairs or ensure producing a feasible solution in the test stage. Other deep learning-based ideas were seen in \cite{joung2016machine, chen2019intelligent, lin2021deep, elbir2019joint} using either supervised learning or unsupervised learning variants, but similar challenges remain.

		\section{Optimal Joint (R)BF\&AS via B\&B}\label{sec:bb}
		A natural idea for solving hard optimization problems is to employ a {\it global optimization} technique, e.g., the B\&B procedure \cite{land1960automatic, clausen1999branch, boyd2007branch}. 
		Designing a practically working B\&B algorithm is often an art---it normally involves judicious exploitation of problem structures. 
		That is, not every hard problem enjoys an efficient B\&B algorithm.
		Nonetheless, as we will see, the special properties of BF and RBF allows for an effective B\&B design. 
		
		\subsection{Preliminaries of B\&B}
		{\color{black}
			We follow the notations from the tutorial in \cite{boyd2007branch} to give a brief {overview of B\&B's design principles}. Consider a nonconvex problem:
			\begin{subequations}\label{eq:bb}
				\begin{align}
					\minimize_{\x}&~f(\x)\\
					{\rm subject~to}&~\x \in {\cal X}.
				\end{align}
			\end{subequations}
			where both the objective function and the constraint can be nonconvex. 
			Suppose that there is a partition of the space ${\cal X}={\cal X}_1\cup \ldots \cup {\cal X}_S$, and that lower and upper bounds of $f(\x)$ over each ${\cal X}_i$ are easier to find (relative to directly solving \eqref{eq:bb}).
			{Let $\Phi_{\rm lb}(\cX_i)$ and $\Phi_{\rm ub}(\cX_i)$ be the algorithms that return lower and upper bounds of the optimal solution of \eqref{eq:bb} over the set $\cX_i$, respectively. Then, the following holds: 
				\begin{align}\label{eq:bbepsilon_1}
					\min_{1\leq i\leq S} \Phi_{\rm lb}({\cal X}_i)  \leq \Phi({\cal X}) \leq    \min_{1\leq i\leq S} \Phi_{\rm ub}({\cal X}_i).
				\end{align}
				where $\Phi(\cX)$ represents the optimal solution of \eqref{eq:bb} over the feasible region $\cX$.}
			A premise of the success of B\&B is that one could find a partition ${\cal X}_i$ for $i=1,\ldots,S$ and a pair of functions $\Phi_{\rm lb}$ and $\Phi_{\rm ub}$ which can make the following hold: 
			\begin{equation}\label{eq:stopcriterion}
				\min_{1\leq i\leq S} \Phi_{\rm ub}({\cal X}_i)  -  \min_{1\leq i\leq S} \Phi_{\rm lb}({\cal X}_i)\leq \epsilon
			\end{equation}
			where $\epsilon > 0$ is a pre-specified error tolerance parameter.

			The effectiveness of B\&B relies on two key factors. First, the design of the lower and upper bounding algorithms represented by $\Phi_{\rm lb}(\cX_i)$ and $\Phi_{\rm ub}(\cX_i)$, respectively, plays a central role. Second, the way of partitioning the space ${\cal X}$ also matters. It often requires a problem-specific way to progressively and judiciously partition the constraint set ${\cal X}$ (usually from rough to fine-grid), so that the difference in \eqref{eq:stopcriterion} could shrink quicker than exhaustive search. {Meeting either of the design requirements is not necessarily easy. Moreover, the key designs in B\&B algorithms (e.g., the ${\cal X}$ partition strategies) are highly problem-dependent; that is, there is hardly a ``standard recipe'' for B\&B algorithm design.}
		}

		\subsection{Proposed B\&B for Joint (R)BF\&AS}
		Problem \eqref{eq:metaproblem} involves optimization in discrete and continuous spaces in the constraints. Designing a B\&B algorithm for such problems can be difficult due to the large search space that consists of both types of constraints. 
		However, the special structure of (R)BF in \eqref{eq:metaproblem} allows us to 
		{efficiently obtain bounds over the entire range of the values of the continuous (beamforming) parameters once the discrete (antenna selection) parameters have been chosen; this will be clearer in \eqref{eq:node_problem} and \eqref{eq:relaxation_node}. As such, we only need to construct a B\&B tree over the discrete space.}

		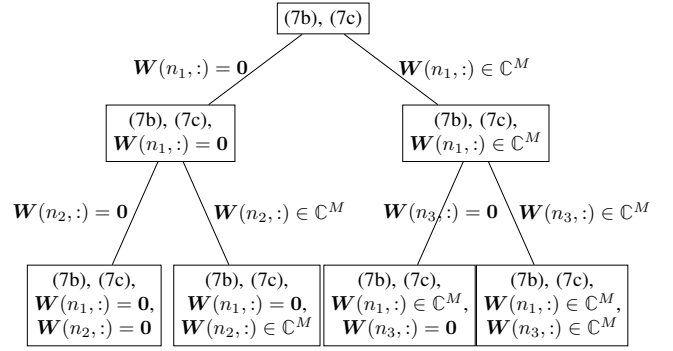
\begin{figure}
			\centering
			\resizebox{\linewidth}{!}{
				\begin{tikzpicture}
					[
					level 1/.style = {level distance = 2cm, sibling distance = 0.6\linewidth},
					level 2/.style = {level distance = 3cm, sibling distance = 0.3\linewidth},
					state/.style = {shape = rectangle, draw, align=center}
					]
					
					\node[state, align=center] {\eqref{eq:sinr_constraint}, \eqref{eq:cardinality}}
					child {node[state, align=center] {\eqref{eq:sinr_constraint}, \eqref{eq:cardinality}, \\ $\W(n_1,:) = \zero$}
						child {node[state, align=center] {\eqref{eq:sinr_constraint}, \eqref{eq:cardinality}, \\ $\W(n_1,:) = \zero$, \\ $\W(n_2,:) = \zero$} 
							edge from parent node [left]{$\W(n_2, :) = \zero$}}
						child {node[state, align=center] {\eqref{eq:sinr_constraint}, \eqref{eq:cardinality}, \\ $\W(n_1,:) = \zero$, \\ $\W(n_2,:) \in \bbC^M$}
							edge from parent node [right]{$\W(n_2,:) \in \bbC^M$}}
						edge from parent node [left]{$\W(n_1, :) = \zero$}
					}
					child {node[state] {\eqref{eq:sinr_constraint}, \eqref{eq:cardinality}, \\ $\W(n_1,:) \in \bbC^{M}$}
						child {node[state, align=center] {\eqref{eq:sinr_constraint}, \eqref{eq:cardinality}, \\ $\W(n_1,:) \in \bbC^M$, \\ $\W(n_3,:) = \zero$} 
							edge from parent node {$\W(n_3, :) = \zero$}}
						child {node[state, align=center] {\eqref{eq:sinr_constraint}, \eqref{eq:cardinality}, \\ $\W(n_1,:) \in \bbC^M$, \\ $\W(n_3,:) \in \bbC^M$}
							edge from parent node [right]{$\W(n_3,:) \in \bbC^M$}}
						edge from parent node [right] {$\W(n_1,:) \in \bbC^M$}
					};
					
				\end{tikzpicture}
			}
			\caption{Illustration of B\&B tree for problem \eqref{eq:metaproblem}. Here $n_i \in [N]$ are the branching variables selected at each node. }
			\label{fig:bb_tree}
		\end{figure}
		
		\subsubsection{B\&B Tree Construction}\label{sec:proposed_bb}  We illustrate the idea of systematically partitioning the feasible region of
		Problem~\eqref{eq:metaproblem} in Fig.~\ref{fig:bb_tree}.
		Here, ${\cal N}_i^{(\ell)}$ denotes the {feasible region corresponding to} the $i$th node at the $\ell$th level. {In the sequel, we will use the term ``node'' and the associated feasible region interchangeably}. The root is denoted as ${\cal N}^{(0)}$, and we have
		\[  {\cal N}^{(0)} = \{  \W~|~\W~\text{satisfies}~\eqref{eq:sinr_constraint},\eqref{eq:cardinality} \}. \]
		In the first level, the region represented by the root node is split into two regions represented by two child nodes, namely,
		\begin{align*}
			{\cal N}_1^{(1)} &= \{  \W~|~\W(n_1,:)=\bm 0,~\W~\text{satisfies}~\eqref{eq:sinr_constraint},\eqref{eq:cardinality} \}\\
			{\cal N}_2^{(1)} &= \{  \W~|~\W(n_1,:) \in \bbC^{M},~\W~\text{satisfies}~\eqref{eq:sinr_constraint},\eqref{eq:cardinality} \}.
		\end{align*}
		{where $n_1 \in [N]$ is an antenna index selected by a designed antenna selection} criterion (e.g., via random sampling). 
		Up to the first level of the tree, the status (``include (activate)'' or ``exclude (shut down)'') of all antennas other than antenna $n_1$ have not been decided. 
		
		Note that the nodes in the B\&B tree could constitute a partition in various forms.
		For example, for nodes in the same level, we have 
		\[ {\cal N}_1^{(\ell)} \cup \ldots \cup {\cal N}^{(\ell)}_{S_\ell} = {\cal N}^{(0)}, \]
		where ${S}_\ell = 2^{\ell}$ is the number of nodes in the $\ell$th level of the tree. In addition, we have
		\begin{align}\label{eq:child_node_generation}
			{\cal N}^{(\ell)}_s = {\cal N}^{(\ell+1)}_{s_1}\cup {\cal N}^{(\ell+1)}_{s_2}, 
		\end{align}
		where $s_1 := 2(s-1) +1$ and $s_2 := 2(s-1)+2$ represent the left and right children developed from ${\cal N}^{(\ell)}_s$ in the full tree.
		In fact, the children of $ {\cal N}^{(\ell)}_s$ in any level and ${\cal N}^{(\ell)}_{-s}$ also present a partition of the root node, where ${\cal N}^{(\ell)}_{-s}$ is the union of ${\cal N}^{(\ell)}_1,\ldots,{\cal N}^{(\ell)}_{S_\ell}$ with ${\cal N}^{(\ell)}_s$ excluded. 
		
		The B\&B algorithm starts from the first level to compute lower and upper bounds of \eqref{eq:metaproblem} over the node-defined regions.
		Then, the B\&B algorithm picks a node to ``branch'', i.e., to further partition oftentimes using a heuristic-based metric; see \cite{clausen1999branch}.
		Going deeper in the tree towards the final leaves will allow us to progressively decide which antennas to activate or shut off.
		Let $t$ denote the iteration index of the B\&B algorithm, where an iteration corresponds to a branching (partitioning a node) operation.
		{Use ${\cal P}^{(t)}$ to denote the collection of $(s,\ell)$ corresponding to the unbranched nodes. Then, the union of ${\cal N}_{s}^{(\ell)}$'s for $(s,\ell) \in {\cal P}^{(t)}$ represents a partitioning of the root in iteration $t$.}
		In each iteration $t$, the stopping criterion in \eqref{eq:stopcriterion} is evaluated. It follows that the following two quantities need to be evaluated: 
		\[ l_G^{(t)} = \min_{(s, \ell) \in {\cal P}^{(t)}}\Phi_{\rm lb}({\cal N}_s^{(\ell)}),~ u_G^{(t)} = \min_{(s, \ell) \in {\cal P}^{(t)}}\Phi_{\rm ub}({\cal N}_s^{(\ell)}), \]
		where $l_G^{(t)}$ and $u_G^{(t)}$ are the global lower and upper bounds in iteration $t$.
		In particular, the lower and upper bounds over the {\it newly} created two child nodes need to be found---since other nodes have been evaluated in a certain previous iteration. The hope is that one would not need to visit all nodes of tree before reaching the stopping criterion in \eqref{eq:stopcriterion}.
		
		\subsubsection{Lower and Upper Bounds} 
		In order to compute $\Phi_{\rm lb}({\cal N}_s^{(\ell)})$ and $\Phi_{\rm ub}({\cal N}_s^{(\ell)})$, let us define $\cA^{(\ell)}_s \subseteq [N]$ and $\cB^{(\ell)}_s \subseteq [N] \backslash \cA_s^{(\ell)}$ to be the index sets of the antennas that have been activated and shut down at node $s$ in level $\ell$, respectively. {Note that $\cA_s^{(\ell)} \cup \cB_s^{(\ell)}\subseteq [N]$ constitute the set of decided antennas at the node.} 
		Then, finding the upper and lower bounds of $\|\W\|_F^2$ at this node amounts to finding those of the following optimization problem:
		\begin{align}  \label{eq:node_problem}
			\minimize_{\W} ~ & \|\W\|_F^2  \\
			\text{subject to } ~ & \cC(\w_m, \h_m, \varepsilon_m, \sigma_m) \geq \gamma_m,~\forall m, \nonumber \\
			& \W(n, : ) = \zero, \quad\forall n \in {\cal B}^{(\ell)}_s, \nonumber \\
			& \W(n, :) \in \bbC^M , \quad \forall n \in {\cal A}^{(\ell)}_s, \nonumber \\
			& \|\W\|_{\rm row-0} \leq L, \quad n \in [N]. \nonumber
		\end{align}
		For any given node ${\cal N}_s^{(\ell)}$, the lower bound can be obtained by solving the following relaxation of \eqref{eq:node_problem}:
		\begin{subequations} \label{eq:relaxation_node}
			\begin{align}
				\Phi_{\rm lb}({\cal N}_s^{(\ell)}) =& \minimize_{\W} ~ \|\W\|_F^2 \\
				\text{subject to } ~ & \cC(\w_m, \h_m, \varepsilon_m, \sigma_m) \geq \gamma_m, \forall m, \\
				& \W(n,:) = \zero, \quad\forall n \in \cB_s^{(\ell)}, \nonumber
			\end{align}
		\end{subequations}
		where we have dropped $\|\W\|_{\rm row-0} \leq L$. If Problem \eqref{eq:relaxation_node} is not feasible, $\Phi_{\rm lb}(\cN_s^{(\ell)})$ is set to {$+\infty$}.
		
		In the following lemma,  we show that \eqref{eq:relaxation_node} can be optimally solved for all nodes in the B\&B tree. It also helps derive a procedure for $\Phi_{\rm ub}(\cdot)$.
		\begin{lemma}\label{lem:solvinglowerbound}
			Regarding \eqref{eq:relaxation_node},  the following hold:
			\begin{itemize}
				\item[(a)] 
				Consider the BF case where perfect CSI is given. Then, \eqref{eq:relaxation_node} can be optimally solved by using SOCP.
				\item[(b)] 
				Consider the RBF case where imperfect CSI is given.  Assume that
				\begin{align}\label{eq:cond_rbf_subset}
					\frac{\|\bm \Pi_{m} {\widetilde{\h}}_m\|^2_2}{\varepsilon^2_m} > 1 + M + (M - \frac{1}{M}) \gamma_m, \forall m ,    
				\end{align}
				where ${\bm \Pi}_{m} := \I - {\widetilde{\H}}_{-m} ({\widetilde{\H}}_{-m}^H {\widetilde{\H}}_{-m})^{-1} {\widetilde{\H}}_{-m}^H $, holds for $\widetilde{\H} \in \{{\H}({\cS},:) | \forall \cS \in [N], |\cS|\geq L\}$. 
				Then, Problem~\eqref{eq:relaxation_node} can be optimally solved using SDR.
				\item[(c)] Under the same conditions of (a) and (b), solving the following gives a valid upper bound of \eqref{eq:node_problem} under the BF and RBF cases, respectively:
				\begin{subequations} \label{eq:upper_bound}
					\begin{align}
						\Phi_{\rm ub}({\cal N}_s^{(\ell)}) &= \minimize_{\W} ~ \|\W\|_F^2 \\
						\text{subject to } &~ \cC(\w_m, \h_m, \varepsilon_m, \sigma_m) \geq \gamma_m, \forall m, \\
						& \W(n,:) = \zero, \quad\forall n \in \widetilde{\cB}_s^{(\ell)}, \nonumber
					\end{align}
				\end{subequations}
				where $\widetilde{\cB}_s^{(\ell)} = \cC_s^{(\ell)} \cup \cB_s^{(\ell)}$ represents the set of $N-L$ antennas to be excluded, and $\cC_s^{(\ell)} \subseteq [N] \backslash (\cA_s^{(\ell)} \cup \cB_s^{(\ell)})$ is the index set of undecided antennas that have been assigned the minimum power in the solution of \eqref{eq:relaxation_node}. If Problem \eqref{eq:upper_bound} is not feasible, $\Phi_{\rm ub}(\cN_s^{(\ell)})$ is {notationally set to $+\infty$.}

			\end{itemize}
		\end{lemma}
		The proof of Lemma~\ref{lem:solvinglowerbound} is relegated to Appendix \ref{app:proof_lemmas}.

		\subsubsection{Node Selection and Branching} 
		After \eqref{eq:relaxation_node} and \eqref{eq:upper_bound} are computed in iteration $t$, $l_G^{(t+1)}$ and $u_G^{(t+1)}$ are updated.
		If the stopping criterion $u_G^{(t)}-l_G^{(t)} \leq \varepsilon$ is not met, one needs to pick a node in ${\cal P}^{(t)}$ to further partition. To this end, we employ the ``lowest lower bound first'' principle that is often used in the literature \cite{clausen1999branch}.  To be specific, we pick a non-leaf node ${\cal N}_{s^\star}^{(\ell^\star)}$ such that
		\begin{align}\label{eq:node_selection}
			(\ell^\star,s^\star)\in \arg\min_{(s,\ell)\in {\cal P}^{(t)} \backslash \cS_{\rm leaf}}~\Phi_{\rm lb}({\cal N}_s^{(\ell)}),
		\end{align}
		where $\cS_{\rm leaf} := \{(\ell, s): |\cA_s^{(\ell)}| = L, |\cB_s^{(\ell)}| = N - L \}$ is the set of leaf nodes. 
		To partition the region ${\cal N}_{s^\star}^{(\ell^\star)}$, we need to pick an {\it undecided} antenna and decide whether to include or exclude it in our solution.
		We select the antenna that has been assigned the largest power among the undecided antennas in iteration $t$, i.e.,
		\begin{align} \label{eq:variable_selection}
			n^\star = \arg \max_{i \in [N] \backslash (\cA_{s^\star}^{(\ell^\star)} \cup \cB_{s^\star}^{(\ell^\star)})} \| \W_{s^\star}^{(\ell^\star)}(i, :) \|_2^2,
		\end{align}
		where $\W_{s^\star}^{(\ell^\star)} :=\arg\min_{\W} \eqref{eq:relaxation_node}$ at ${\cal N}_{s^\star}^{(\ell^\star)}$.
		Then, $n^\star$ is used to partition ${\cal N}_{s^\star}^{(\ell^\star)}$ into two child nodes (i.e., excluding and including antenna $n^\star$ on top of the decided antennas in ${\cal N}_{s^\star}^{(\ell^\star)}$). 
		{The associated include/exclude sets in the child nodes, $\cN_{s^\star_i}^{(\ell^\star + 1)}, i \in \{1,2\}$, are updated as follows:
			\begin{align*}
				\cB_{s^\star_1}^{(\ell + 1)} &= \cB_{s^{\star}}^{(\ell)} \cup \{n^\star\},  \quad \cA_{s^\star_1}^{(\ell +1)} = \cA_{s^\star}^{(\ell)}, \\
				\cA_{s^\star_2}^{(\ell + 1)} &= \cA_{s^{\star}}^{(\ell)} \cup \{n^\star\},  \quad \cB_{s^\star_2}^{(\ell +1)} = \cB_{s^\star}^{(\ell)}.
		\end{align*}}
		Note that if any of the child nodes, have $L$ included or $N-L$ excluded antennas, we apply the following update:
		\begin{align}\label{eq:determine_node}
			\cB_{s^\star_i}^{(\ell^\star +1)} &= [N] \backslash \cA_{s^\star_i}^{(\ell^\star +1)} \text{ if } |\cA_{s^\star_i}^{(\ell^\star +1)}| = L \nonumber\\
			\cA_{s^\star_i}^{(\ell^\star +1)} &= [N] \backslash \cB_{s^\star_i}^{(\ell^\star +1)} \text{ if } |\cB_{s^\star_i}^{(\ell^\star +1)}| = N-L. 
		\end{align}
		This ensures that we do not generate any new nodes that do not satisfy \eqref{eq:cardinality}.
		Finally, the two children replace ${\cal N}_{s^\star}^{(\ell^\star)}$ in ${\cal P}^{(t)}$ to form ${\cal P}^{(t+1)}$.

		Note during the process, some nodes in the B\&B tree can be simply discarded, or, ``fathomed''---as in the standard terminologies of B\&B \cite{clausen1999branch}. After iteration $t$, one can potentially find a set of $(s',\ell')$ such that
		\[ \Phi_{\rm lb}({\cal N}_{s'}^{(\ell')}) > u_G^{(t)}. \]
		The above means that ${\cal N}_{s'}^{(\ell')}$ needs not to be further partitioned in the next iteration. Hence, we can form a set $\cF^{(t)}$ in each iteration, which only contains the nodes that need to be further considered, i.e., 
		$$ \cF^{(t)} = \left\{ \left(s',\ell'\right) \in {\cal P}^{(t)}~\left|~ \Phi_{\rm lb}\left({\cal N}_{s'}^{(\ell')}\right) {\leq} u_G^{(t)} \right. \right\}$$
		This is arguably the most important for attaining efficiency against exhaustive search.  A summary of the B\&B procedure can be found in Appendix~\ref{app:bb_procedure}.
		
		\subsubsection{An Alternative B\&B Method}\label{remark:alternative_formulation}
		It is interesting to note that there is often more than one way to come up with a B\&B procedure for a given problem.
		For example,
		a commonly used approach for deriving B\&B of {\it mixed integer and linear programs} (MILPs), and more generally, subset selection problems (see, e.g., \cite{bertsimas2016best, lee2019learning}) can also be used for our problem \eqref{eq:metaproblem}.
		The method is by introducing auxiliary Boolean variables.
		Specifically, problem~\eqref{eq:metaproblem} can be expressed as follows:
		\begin{subequations}\label{eq:z_formulation}
			\begin{align}
				\minimize_{\W, \z} &~\|\bm W\|_F^2 \\
				{\rm subject~to}&~{\cal C}(\bm w_m,\bm h_m, \varepsilon_m, \sigma_m) \geq \gamma_m,  \nonumber \\
				&~ \z \in \{0, 1\}^N \label{eq:z01},  \\
				&~ \z^\T \one \leq L, \nonumber\\
				&~ \|\W(n,:)\|_2 \leq C z(n), ~\forall n \in [N]. \nonumber
			\end{align}
		\end{subequations}
		where $C< \infty$ is a large positive constant and $z(n) = 0$ means that the $n$th antenna is excluded whereas $z(n) = 1$ indicates the opposite.
		The constraint in \eqref{eq:z01} can be relaxed to be $\z\in[0,1]^N$ for finding the lower bound (see Appendix \ref{app:alternative_formulation} for details). 
		In this procedure, the branching operations are imposed on the new variable $\z$ \cite{bertsimas2016best, lee2019learning}. {The reason that we do not choose formulation ~\eqref{eq:z_formulation} to design B\&B for our joint (R)BF\&AS problem is that this approach could be computationally (much) less efficient compared to the proposed approach (see a proof in Theorem \ref{thm:bb_convergence} in the next subsection}).
		{The computational efficiency of our method comes from the fact that the computation of upper and lower bounds in \eqref{eq:relaxation_node} and \eqref{eq:upper_bound} can be reused for many nodes; see the proof of Theorem~\ref{thm:bb_convergence}. However, it is not obvious if such kind of computation reduction is still possible for the formulation in~\eqref{eq:z_formulation}.}

		\subsection{Optimality}
		We show that the proposed algorithm will produce optimal solutions for the problem of interest:

		\begin{theorem}\label{thm:bb_convergence}
			Regarding the proposed B\&B procedure (see Appendix~\ref{app:bb_procedure}), the following statements hold:
			\begin{itemize}
				\item[(a)] When BF is considered, the proposed B\&B solves \eqref{eq:metaproblem} optimally.
				\item[(b)] When RBF is considered, if the conditions in Lemma \ref{lem:solvinglowerbound}(b) are satisfied, the proposed B\&B solves \eqref{eq:metaproblem} optimally.
				\item[(c)] The total number of SOCPs/SDRs {solved} by the proposed B\&B is upper bounded by 
				$$ Q_{\rm Compute} = {N \choose L} + \sum_{i=2}^{N-L + 1} {N-i \choose L-1}.$$
				The number of SOCPs/SDRs needed by the B\&B associated with the alternative formulation in Sec.~\ref{remark:alternative_formulation} is upper bounded by $Q_{\rm Compute}'=2{N \choose L} - 1$. 
			\end{itemize}
			
		\end{theorem}
		
		\color{black}
		The proof of Theorem \ref{thm:bb_convergence} is in Appendix \ref{app:proof_bb}.
		At the first glance, it feels a bit surprising that the B\&B algorithms could use more than ${N\choose L}$ SOCP/SDRs to find the optimal solution, since this seems to be worse than exhaustive search. This is because, in the worst case, B\&B visits many more intermediate states in the search tree---but exhaustive search only visits the leaves. Nonetheless, in practice, B\&B is often much more efficient than exhaustive search since B\&B does not really exhaust all the nodes. 
		Theorem~\ref{thm:bb_convergence} (c) spells out the advantage of our B\&B design relative to the more classic B\&B idea as in \eqref{eq:z_formulation} from the MILP literature.
		Note that the reduction of complexity shown in (c) could be substantial.
		For example, when $(N,L) = (12,8)$, $Q_{\rm Compute}=$660, whereas $Q_{\rm Compute}'=$989. Hence, there is a potential saving of 339 SOCPs/SDRs (reduction by 34\%) in the worst case.
		
		\begin{remark}
			Under approximate CSI, the conditions in Lemma \ref{lem:solvinglowerbound}(b) is the premise for our theorem to hold [cf. Theorem~\ref{thm:bb_convergence}(b)].
			When the condition is violated, {it is possible that the SDR in \eqref{eq:node_problem} might} return solutions whose rank is higher than one {\it in theory}---which would hinder the optimality of the B\&B procedure. Nonetheless, such higher-rank solutions were never seen in our simulations---which is consistent with observations from the literature \cite{ma2017unraveling, song2012robust, chang2011worst, zheng2008robust}.
			Our conjecture is that the sufficient condition in Lemma \ref{lem:solvinglowerbound}(b) is far from necessary.
			In {rare} cases where rank-one solutions do not exist for \eqref{eq:node_problem}, standard procedures like randomization \cite{luo2010semi} may be resorted to for finding rank-one approximations.
		\end{remark}

		\section{Accelerated Joint (R)BF\&AS via ML}
		The challenge of any B\&B algorithm lies in the large number of nodes in the tree. This means that in the worst case, many SOCPs and SDRs need to be solved.
		An idea from the ML community is to ``train'' a classifier to recognize the {\it relevant nodes}, i.e., nodes that lead to leaves containing the optimal solution \cite{he2014learning}. If a node is deemed to be ``irrelevant'', the B\&B algorithm would simply skip branching on this node, and thus could save a substantial amount of time.
		In this section, we will show that a similar idea can be used for accelerating our B\&B based joint (R)BF\&AS algorithm---with carefully designed neural models to meet the requirements arising in wireless communications. More importantly, we will present comprehensive performance characterizations, including sample complexity and global optimality retention, which are currently lacking in the existing literature.

		\subsection{Preliminaries: Node Classification and Imitation Learning}\label{sec:ml}
		\subsubsection{Node Classification}
		Let us denote $$\bpi_{\btheta}: \bbR^{P} \to [0, 1]$$ as the {node} classifier parameterized by $\bm \theta$, which returns the probability of a node being relevant. Let $$\bm \phi( \cN^{(\ell)}_s) \in \bbR^P$$ be the mapping from a node to its feature representation. 
		When $\bpi_{\btheta}(\bm \phi(\cN^{(\ell)}_s)) < 0.5$, then the node is deemed irrelevant. Otherwise, the node is branched.

		To train such a classifier, denote
		$\{ ({\cal N}_s, y_s) \}_{s=1}^T$ as the (node, label) training data, where we have removed the level indices of the nodes for notation simplicity. 
		To create the training pairs, one could run random problem instances of \eqref{eq:metaproblem} using the B\&B procedure.
		Note that the label $y_s$ is annotated according to the following rule:
		\begin{align}
			y_s=\begin{cases} 1,\quad {\cal A}_s \subseteq {\cal A}^\star\text{~and~}\cB_s \subseteq [N] \backslash \cA^\star,\\
				0,\quad{\rm otherwise},
			\end{cases}
		\end{align}
		where {$\cA_s$ and $\cB_s$ are the index sets of included and excluded antennas at node $s$, respectively, and} $\cA^\star$ is the index set of the active antennas of the optimal solution found by B\&B of the associated problem instance.

		\subsubsection{Imitation Learning}\label{sec:imitation_learning}
		The simplest supervised learning paradigm would learn $\bpi_{\btheta}$ using the following risk minimization criterion:
		\begin{equation}\label{eq:supervised}
			\minimize_{\btheta}~\frac{1}{T}\sum_{s=1}^T{\cal L}\left( \bpi_{\btheta}\left(\bm \phi_s\right),y_s\right) + r(\bm \theta),
		\end{equation}
		where $\bm \phi_s := \bm \phi(\cN_s)$, ${\cal L}(x,y)$ is a certain loss function, e.g., the logistic loss, and $r(\bm \theta)$ is a regularization term, e.g., $r(\btheta)=\lambda\|\btheta\|_2^2$.
		Unfortunately, such a supervised learning approach often does not work well, since it ignores the fact that the node generating process is {\it sequential} and {\it interactive} with the node classifier in the test stage. 
		In ML-based MILP, the remedy is to {adopt the {\it imitation learning} (IL) \cite{ross2011reduction} approach, where $\bpi_{\btheta}$ is integrated in the training data generating process \cite{he2014learning}.}
		To be more specific, the training data generation process is done in a batch-by-batch manner with {\it online optimization}.
		The IL training criterion is as follows {(see Section \ref{sec:data_gen} for data generation and training process)}: 
		\begin{align}\label{eq:online}
			\btheta^{(i+1)} &= \\
			&\arg \min_{\btheta} \frac{1}{i} \sum_{t=1}^i \frac{1}{|\cD_t|}\sum_{(\bm \phi_{s},y_{s}) \in \cD_t} {\cal L}\left(\bpi_{\btheta}(\bm \phi_{s}), y_{s} \right) + r(\bm \theta),\nonumber
		\end{align}
		where $\cD_t$ is the $t$th batch of training pairs. The learned model parameter $\widehat{\btheta}$ is selected from $\btheta^{(i)}$'s via the following:
		\begin{align}\label{eq:theta_selection}
			\widehat{\btheta} = \arg \min_{\btheta \in \{ \btheta^{(i)}\}_{i=1}^I} \bbE_{(\bm \phi_{s},y_{s})} \left[{\cal L}\left(\bpi_{\btheta}(\bm \phi_{s}), y_{s} \right)\right],
		\end{align}
		where $I$ is the total number of batches generated during the training process.
		In practice, one can use a validation set to approximate the above expectation. 
		In the test stage, the proposed B\&B algorithm is run with the assistance of $\bpi_{\widehat{\btheta}}$.

		\smallskip
		
		The key of using IL to accelerate the proposed B\&B for joint (R)BF\&AS is twofold, namely, a practical node classifier tailored for wireless communications and a convergent online training algorithm. We will detail our designs to address the two requirements in the next subsections.

		\subsection{GNN-based Node Classifier for Joint (R)BF\&AS}\label{sec:gnn_architecture}
		To design the node classifier, a critical consideration in wireless communications is that the number of users to serve could drastically change from time to time. This requires us to design an ML model that is agnostic to such changes, as re-training a model when change happens is not affordable. 
		Towards this end, we design a GNN-based node classifier \cite{scarselli2008graph}. Note that GNNs learn aggregation operators over a graph, and thus is naturally robust to the change of entities on the graph. We will leverage this property to design our node classifier.

		\begin{figure}
			\centering
			\includegraphics[width=0.4\linewidth]{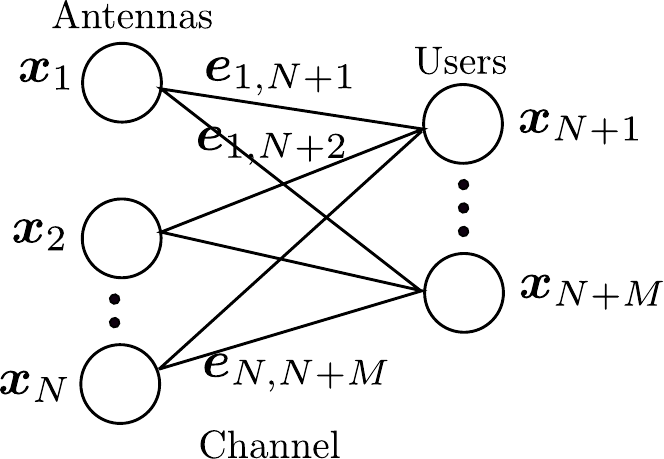}
			\caption{Illustration of the input graph representation for a node.}
			\label{fig:gnn_feature_illus}
		\end{figure}

		To describe the GNN-based node classifier, we first define a graph to represent ${\cal N}_s^{(\ell)}$. Fig. \ref{fig:gnn_feature_illus} illustrates the idea, where the antennas and users represent the vertices, and the channel represent the edge between the vertices. It is important to design the features of the vertices and the edges, so that they represent the essential information of the node ${\cal N}_s^{(\ell)}$. To be specific, we let
		\begin{align}\label{eq:Va}
			& \x_n \in \mathbb{R}^{V_a}, n \in [N], \quad \x_{N+m}\in \mathbb{R}^{V_u}, m \in [M] , \text{ and} \nonumber \\ 
			& \e_{n, N+m} \in \mathbb{R}^{V_e}, n \in [N], m \in [M]
		\end{align}
		represent the feature vectors of antenna $n$ (a vertex),
		user $m$ (a vertex), and the channel between the antenna $n$ and the user $m$ (an edge), respectively.  Layer $d$ of the GNN ``aggregates'' the embedding of graph neighbors to update the $u$th vertex for all $u \in [M+N]$.
		The definition of such aggregation can be flexible. For example, in the
		\textit{message passing neural network} \cite{gilmer2017neural}, the aggregation is done by the following:
		\begin{align}\label{eq:gnn_node_update}
			\q_u^{(d)} = \bm \xi (\Z_1 \q_u^{(d-1)} + \sum_{v \in \cE_u} \bm \xi (\Z_2 \q_v^{(d-1)} + \Z_3 \e_{u,v} ) ),
		\end{align}
		where $\bm q_u^{(0)}=\bm x_u$; $\bm Z_i$ for $i=1,2,3$ are the {\it aggregation operators} of the GNN; $\bm \xi(\cdot)$ represents the activation functions of layer $d$; and ${\cal E}_u$ is the index set of all the one-hop neighbors of vertex $u$ on the graph. 
		The output of the GNN is
		$$ \bpi_{\btheta}(\bm \phi_s) = \frac{1}{U} \sum_{u \in [U]} \bm \zeta \left(\bm \beta^\T \q_u^{(D)} \right),~\bm \phi_s =\bm \phi({\cal N}_s)\in\mathbb{R}^P$$
		where $U=M+N$ is the total number of vertices; $\bm \phi (\cN_s) = [\x_1^\T, \dots,\x_{N+M}^\T, \e_{1,N+1}^\T, \dots, \e_{N,N+M}^\T]^\T$; and $\bm \zeta(\cdot)$ is a sigmoid function. 
		Here, the parameter to be optimized is given by $\btheta := [{\rm vec}(\Z_1)^\T, {\rm vec}(\Z_2)^\T, {\rm vec}(\Z_3)^\T, \bm \beta^\T ]^\T$.

		Table~\ref{tab:feature_design} shows the detailed feature descriptions. 
		We design two types of features to represent the B\&B nodes.
		To be specific, Type I features represent the features whose dimensions are not affected by the problem size parameters $N,M,L$. For example, $\Phi_{\rm lb}$ is a Type I feature as it is always a scalar under any $(N,M,L)$. Type II features are those whose dimensions change when $(N,M,L)$ changes. For instance, the channel matrix $\H\in\mathbb{C}^{M\times N}$ is a Type II feature.

		Appendix~ \ref{sec:app_feature_implementation} details the conversion from the features in Table~\ref{tab:feature_design} to $\x_n$ and $\e_{u,v}$. Note that the special structure of GNN allows us to employ both Type I and Type II features. The reason is that
		the change of $M,N$ and $L$ only changes the number of vertices/edges of the graph in Fig.~\ref{fig:gnn_feature_illus}. This does not necessarily change $V_a$, $V_e$ and $V_u$ that determines the size of $\Z_i$ [cf. Eq.~\eqref{eq:Va}]---if $\x_n$ and and $\e_{n,m}$ are designed properly under the GNN framework (see Appendix~ \ref{sec:app_feature_implementation}).
		However, if one uses SVM as in \cite{he2014learning} or other types of neural networks (e.g., fully connected network (FCN) and convolutional neural network (CNN)), Type II features are much less flexible to use.
		We should remark that our feature design is not ``optimal'' in any sense,
		but using Type II features arguably provides more comprehensive information about the node and could often enhance the node classification accuracy. 
		
		Table \ref{tab:svm_gnn_comparison} shows numerical evidence to support our postulate. 
		There, different classifiers are trained by IL using problem instances as described in Sec.~\ref{sec:exp}. The FCN has two hidden layers with 32  hidden units in each layer, a sigmoid activation function on the output layer, and ReLU activations on the remaining layers. The architecture of the GNN is described in Appendix~\ref{app:gnn}. 
		The SVM and FCN could only use the Type I features. The GNN with both types of features clearly offers a lower node classification error.

		\begin{remark}\label{rmk:gnn}
			In addition to being able to work with both types of features,
			another important benefit of using GNN is as folows:
			Since $\bm \theta$ of the GNN model does not depend on $(N,M,L)$, the learned model can naturally work when the numbers of users and antennas change, {as long as $V_a$, $V_u$, and $V_e$ remain the same.} 
			That is, the model trained on problem instances with $(N,M,L)$ can be seamlessly tested on cases with $(N',M',L')\neq (N,M,L)$.
			This property of GNN will be vital for applying the proposed method in real-world scenarios where the problem size changes constantly (as the number of users to be served by a BS changes all the time).
			It also helpd scale up the proposed method for coping with large $(N,M,L)$ using a $\bm \theta$ trained from small problem sizes, which could save a substantial amount of computational resources. 
			
			{We should emphasize that GNN is ``insenstive'' to the change of problem size across training and testing. However, drastic change of other aspects (e.g., channel model and noise level) across the two stages does affect the performance more substantially. In other words, beyond the problem size, our GNN-based method still expects that the training and testing data to share similar characteristics, as other machine learning models do.}
			
		\end{remark}

		\begin{table}[]
			\centering
			\caption{Feature Design for the GNN based node classifier. }
			\begin{tabular}{|c|c|}
				\hline
				\textbf{Type I Features} & \textbf{Type II Features}                                 \\ \hline
				$l_G^{(t)}$                                                          & $\cA_s^{(\ell)}$                                                  \\ \hline
				$u_G^{(t)}$                                                          & $\cB_s^{(\ell)}$                                                  \\ \hline
				$\Phi_{\rm lb}(\cN_s^{(\ell)})$                                      & $[\|\W_{\ell, s}(1,:)\|_2^2, \dots, \|\W_{\ell, s}(N,:)\|_2^2]$   \\ \hline
				$\Phi_{\rm ub}(\cN_s^{(\ell)})$                                     & $\H$                                                              \\ \hline
				$\ell$                                                               & $\W_{\rm incumbent}$ (see Algorithm \ref{algo:abb})               \\ \hline
				$\mathbb{1}(\Phi_{\rm ub}(\cN_s^{(\ell)}) - u_G^{(t)} < \epsilon)$.  & $\W_{\ell,s}$                                                     \\ \hline
				& $|\W_{\ell,s}(:,m)^H \h_m|^2$.                                    \\ \hline
				& Aggregate Interference using $\W_{\ell,s}$.                       \\ \hline
			\end{tabular}
			\label{tab:feature_design}
		\end{table}

		\begin{table}[]
			\centering
			\caption{Classification error (\%) attained by SVM, FCN and GNN based classifier for classifying relevance of the nodes. $\gamma_m = \sigma_m = 1, \varepsilon=0.1$.}
			\begin{tabular}{|c|c|c|c|c|}
				\hline
				& \multicolumn{2}{c|}{Perfect CSI} & \multicolumn{2}{c|}{Approximate CSI} \\ \hline
				Problem sizes & \multirow{2}{*}{(4,3,2)} & \multirow{2}{*}{(8,6,4)} &  \multirow{2}{*}{(4,3,2)} & \multirow{2}{*}{(8,5,4)}   \\
				$(N,M,L)$& & & &\\ \hline
				SVM & 8.49 & 16.67 & 7.17 & 11.67 \\ \hline 
				FCN & \textbf{6.93} & 13.95 & 26.95 & 10.18 \\ \hline
				GNN & 7.26 & \textbf{12.23} & \textbf{6.62} & \textbf{8.49}\\ \hline
			\end{tabular}
			\label{tab:svm_gnn_comparison}
		\end{table}

		\subsection{Data Generation and Online Training}\label{sec:data_gen}
		We use an IL framework to train the GNN, which is summarized in
		Algorithm \ref{algo:prune_learning}. The framework is based on the online learning method in \cite{ross2011reduction}. The work in \cite{ross2011reduction} was proposed for convex learning criteria. Necessary modifications are made in Algorithm~\ref{algo:prune_learning} to accommodate our nonconvex learning problem.

		Algorithm \ref{algo:prune_learning} consists of two steps in each iteration: data collection and classifier improvement. In the $i$th iteration, the accumulated dataset $\cD_i$ is obtained by solving B\&B on $R$ problem instances using the current classifier learned from the previous data batches, $\bpi_{\btheta^{(i)}}$. Then, the classifier is retrained using $\cup_{t=1}^i \cD_i$ and 
		\[  \widehat{\bm \theta}^{(i+1)} =  \arg\min_{\bm \theta \in \bm \varTheta}  ~\g_i (\btheta) + r(\bm \theta)       \]
		where $\bm \varTheta$ specifies the constraints of the GNN parameters [cf. Eq~\eqref{eq:Theta}]; {the loss function $g_i(\cdot)$ is defined as follows:}
		\begin{align}\label{eq:g_def}
			\g_i (\btheta) := \frac{1}{i} \sum_{t=1}^i \frac{1}{|\cD_t|} \sum_{(\bm \phi_s, y_s) \in \cD_t} \cL(\bpi_{\btheta}(\bm \phi_s), y_s);   
		\end{align}
		additionally, we select $\bm r(\btheta) = - \bm \psi^\T \btheta$ in which $\bm \psi$ is sampled from exponential distribution in each iteration.
		This specific choice of $\bm r(\bm \theta)$ plays an important role in our nonconvex learning problem (where the nonconvexity arises due to the use of GNN). 
		To be more specific, such a random perturbation-based $r(\bm \theta)$ is advocated by recent developments from nonconvex online learning \cite{agarwal2019learning}. It was shown in \cite{agarwal2019learning} that using $\bm r(\btheta) = - \bm \psi^\T \btheta$ ensures no-regret type convergence of nonconvex online learning. This property is a critical stepping stone towards establishing learning guarantees of our GNN-based framework.
		This will become clearer in the proofs of Theorem~\ref{thm:main}.

		The training data generation subroutine is given in Algorithm \ref{algo:collect_data}. To generate ${\cal D}_i$, the algorithm first runs B\&B on a given problem instance to find the optimal solution. Next, B\&B is run again but with $\bm \pi_{\bm \theta^{(i)}}$ to generate nodes. {The training pairs $(\bm \phi_s, y_s)$} are annotated by utilizing the optimal solution obtained in the first run.

		\begin{algorithm}[t!]\label{algo:prune_learning}
			\footnotesize
			\SetAlgoLined
			Input: $I, R (\text{number of training instances per batch}), \eta$; \\
			$\cD_1 = \{ \}$; \\
			\For{$i = 1$ to $I$ }{
				Sample $\bm \psi \sim ({\rm Exp}(\eta))^B$ \tcp{${\rm Exp}(\eta)$ is the exponential distribution with pdf $p(x) = \eta \exp(-\eta x)$; $\btheta^{(i)}\in\mathbb{R}^B$;}  \label{line:reg}
				\For{$r=1$ to $R$   }{
					Generate problem instance $\sf Q $; \\
					\eIf{i=1}{
						$\cD^{(\sf Q)} \leftarrow $ run $\texttt{BB}({\sf Q})$ and label the nodes using optimal solution; \\
					}{
						$\cD^{(\sf Q)} \leftarrow \texttt{Algorithm\_\ref{algo:collect_data}}(\sf Q, \bpi_{\btheta^{(i)}})$; \\	
					}
					$\cD_i \leftarrow \cD_{i} \cup \cD^{(\sf Q)}$; \\
				}
				$\btheta^{(i+1)} = \arg \min_{\btheta \in \bm \varTheta} ~ \frac{1}{i} \sum_{t=1}^i \frac{1}{|\cD_t|}\sum_{(\bm \phi_s, y_s) \in \cD_t }\cL(\bpi_{\btheta}(\bm \phi_s), y_s) - \bm \psi^\T \btheta$
			}
			Return $\widehat{\bm \theta}=\arg \min_{\btheta \in \btheta_{1:I}} \frac{1}{|\cD^{\rm valid}_i|} \sum_{(\bm \phi_s, y_s) \in \cD^{\rm valid}_i}[ \cL( \bpi_{\btheta}(\bm \phi_s), y_s)]$ \tcp{where $\cD^{\rm valid}_i$ validation batch $i$ generated by B\&B with $\bpi_{\btheta^{(i)}}$}
			\caption{\texttt{Online GNN Learning}}
		\end{algorithm}
		
		\begin{algorithm}[t!]\label{algo:collect_data}
			\footnotesize
			\SetAlgoLined
			Input: $\sf Q$, $\bpi_{\btheta}$; \\
			\tcp{optimal solution and optimal selected antenna subset to problem $\sf Q$} 
			$(\W^\star, \cA^\star) = \texttt{BB}(\sf Q);$ (see Algorithm \ref{algo:bb} in Appendix~\ref{app:bb_procedure} for \texttt{BB}) \\
			Execute Line \ref{line:main_init_start} to Line \ref{line:main_init_end} in Algorithm \ref{algo:abb}; \tcp{Initialization}
			$\cD \leftarrow \{ \}$;  \\
			\While{B\&B termination criteria is not met }{
				Execute Line \ref{line:main_body_start} to Line \ref{line:main_body_end} from Algorithm~\ref{algo:abb}; \\
				\eIf{$\cN_{s^\star}^{(\ell^\star)}$ is relevant}{
					$\cD \leftarrow \cD \cup \{ \bm \phi_{s^\star}^{(\ell^\star)}, 0 \}$; \\
				}{
					$\cD \leftarrow \cD \cup \{ \bm \phi_{s^\star}^{(\ell^\star)}, 1\}$; \\
				}
			}
			Return $\cD$; 
			\caption{\texttt{Training Data Generation} }
		\end{algorithm}
		
		\color{black}
		
		\smallskip
		
		The overall GNN-accelerated B\&B procedure is summarized in Algoirthm~\ref{algo:abb}.
		The algorithm is termed as 
		{\textit{MachINe learning-based joInt beaMforming and Antennas seLection} (\texttt{MINIMAL})}
		The node classifier is  used in Line~\ref{line:ML}.

		\begin{algorithm}[t!]\label{algo:abb}
			\footnotesize
			\SetAlgoLined
			Input: Problem instance $(\h_m, \sigma_m, \gamma_m, \varepsilon_m), \forall m$, trained pruning policy $\bpi_{\btheta}$, relative error $\epsilon$; \\
			\tcp{Add the root node first}
			$\cA^{(0)}_1 \leftarrow \{\}, \cB^{(0)}_1 \leftarrow \{\} $; \label{line:main_init_start}\\
			Select node using \eqref{eq:node_selection} for $\cN^{(0)}_1$; \\
			$\W_{\rm incumbent} \leftarrow$ solution to \eqref{eq:upper_bound}; \\
			$l_G^{(0)} \leftarrow \|\W^{(0)}_1\|_F^2$, $u_G^{(0)} \leftarrow \|\W_{\rm incumbent}\|_F^2$; \\
			$\cF^{(0)} \leftarrow \{ (0,1) \}$; \\
			$t \leftarrow 0$; \label{line:main_init_end}\\
			\While{$|\cF^{(t)}|> 0$ and $\nicefrac{\left|u_G^{(t)} - l_G^{(t)}\right|}{l_G^{(t)}} > \epsilon$ 
			}{
				Select a non-leaf node $(\ell^\star, s^\star)$ using \eqref{eq:node_selection}; \label{line:main_body_start} \\
				Remove the selected node $ \cF^{(t)} \leftarrow \cF^{(t)} \backslash \cN^{(\ell^\star)}_{s^\star} $; \\ 
				\If{$\bpi_{\btheta}\left(\bm \phi^{(\ell^\star)}_{s^\star}\right) \geq 0.5$ \label{line:ML} }{
					Select variable $n^\star$ using \eqref{eq:variable_selection};\\
					Generate child nodes $\cN^{(\ell^\star+1)}_{s^\star_1}$ and $\cN^{(\ell^\star+1)}_{s^\star_2}$ using \eqref{eq:child_node_generation} and append to $\cF^{(t)}$; \\
					$k \leftarrow \arg \min_{i \in \{ 1, 2\}} \Phi_{\rm ub}\left(\cN^{(\ell^\star + 1)}_{s^\star_i}\right)$; \\
					\If{$\Phi_{\rm ub}\left(\cN^{(\ell^\star + 1)}_{s^\star_k}\right) \leq u_G^{(t)}$}
					{
						$u_G^{(t+1)} \leftarrow  \Phi_{\rm ub}\left(\cN^{(\ell^\star + 1)}_{s^\star_k}\right)$; \\
						$\W_{\rm incumbent} \leftarrow $ solution to \eqref{eq:upper_bound} for $\cN^{(\ell^\star + 1)}_{s^\star_k}$; \\
					}
					$l_G^{(t+1)} \leftarrow  {\rm min}_{(\ell, s) \in \cF^{(t)}} \Phi_{\rm lb}\left(\cN^{(\ell)}_s\right)$; \\
					
				}
				${\cal F}^{(t+1)} \leftarrow \left\{(s',\ell') \in {\cal F}^{(t)}~|~ \Phi_{\rm lb}\left({\cal N}_{s'}^{(\ell')}\right) \leq u_G^{(t+1)} \right\}; $ \\
				
				$t \leftarrow t + 1$; \label{line:main_body_end}\\
			}
			Return $\W_{\rm incumbent}$; \\
			\caption{Main Algorithm: \texttt{MINIMAL}}
		\end{algorithm}

		\subsection{Performance Characterizations}
		Our goal is to characterize the performance of \texttt{MINIMAL}, e.g., under what conditions (e.g., the amount of training samples and the complexity of the GNN) \texttt{MINIMAL} can accelerate the proposed B\&B without losing its optimality.
		To our best knowledge, such performance characterization have not been provided for ML-based B\&B acceleration, even when the learning problem is convex.
		
		To proceed, we will use the following assumptions:
		
		\begin{assumption}\label{ass:gnn}
			Assume that the following statements about the data features and the GNN in Sec.~\ref{sec:gnn_architecture} hold:
			\begin{itemize}
				\item[(a)] The input features are bounded, i.e., $\|\x_u\|_2, \|\e_{u,v}\|_2 \leq B_{\x}, \forall u,v$. 
				\item[(b)] The activation functions $\bm \xi (\cdot)$ and $\bm \zeta (\cdot)$ are $C_{\bm \xi}$-Lipschitz and $C_{\bm \zeta}$-Lipschitz continuous, respectively. In addition, $\bm \xi(\zero) = \zero$.
				\item[(c)]  Let $\cL: \bbR \times \bbR \to [-B_{\cL}, B_{\cL}] $ be  $C_{\cL}$-Lipschitz in its first argument, i.e.,
				$ |\cL(x,y) - \cL(x',y)| \leq C_{\cL} |x - x'|  .$
				\item[(d)] The parameters of the GNN are bounded; i.e., $\|\Z_i\|_2 \leq B_{\Z}, \forall i \in \{1,2,3\}$ and $\|\bm \beta\|_2 \leq B_{\bm \beta}$. 
			\end{itemize}
			
		\end{assumption}
		
		Let us define the set of parameters $\bm \varTheta$ as follows:
		\begin{align}\label{eq:Theta}
			\bm \varTheta := \big\{ & \btheta =[{\rm vec}(\Z_1)^\T, {\rm vec}(\Z_2)^\T, {\rm vec}(\Z_3)^\T, \bm \beta^\T ]^\T ~\mid~ \nonumber\\
			& \|\Z_i\|_2 \leq B_{\Z}, \bm \beta \leq B_{\bm \beta}, i \in \{1,2,3\} \big\}.
		\end{align}
		Using the above, we first characterize the generalization error of the GNN with the following Lemma:
		
		\begin{lemma}[Generalization Error of GNN] \label{lemma:generalizatio_error}
			Consider a GNN $\bm \pi_{\bm \theta}$ in Sec. \ref{sec:gnn_architecture} and $\cG = \{ \bm \phi_k, y_k\}_{k=1}^K$ of i.i.d. samples. Then, for $\btheta \in \bm \varTheta$, the following holds with probability at least $1-\delta$: 
			\begin{align}\label{eq:generalization_error}
				&{\sf Gap}(\delta, K) \\
				&:=  \bbE[\cL(\bpi_{\btheta}(\bm \phi), y)] - \nicefrac{1}{K} \sum_{(\bm \phi_k, y_k) \in \cG} \cL(\bpi_{\btheta}(\bm \phi_k), y_k)  \nonumber\\
				& \leq \frac{8C_{\cL}}{K} + \frac{24 C_{\cL} B_{\cL}}{\sqrt{K}} \sqrt{(3E^2 + E)\log \Lambda} + 3B_{\cL}\sqrt{\frac{\log{(2/\delta)}}{2K}}, \nonumber
			\end{align}
			where $  \alpha = ((1+UC_{\bm \xi}) C_{\bm \xi} B_{\Z})$,
			\begin{align*}
				\Lambda &= 1 + 12 \sqrt{EK}  B_{\Z} {\rm max} \{\Sigma_{\Z_1},\Sigma_{\Z_2},\Sigma_{\Z_3}, \nicefrac{B_{\bm \beta}}{B_{\Z}}\Sigma_{\bm \beta}\}, \\
				\Sigma_{\Z_1} &= C_{\bm \zeta} B_{\bm \beta} U C_{\bm \xi}^3 B_{\Z} B_{\x} \frac{\alpha^{(D+1)} - 2 \alpha + 1}{(\alpha - 1)^2}, \Sigma_{\Z_2} = U C_{\bm \xi} \Sigma_{\Z_1},\\
				\Sigma_{\Z_3} &= C_{\bm \zeta} B_{\bm \beta} U C_{\bm \xi}^2 B_{\Z} B_{\x} \frac{\alpha^D - 1}{\alpha - 1}, \\
				\Sigma_{\bm \beta} &=  C_{\bm \zeta}  B_{\x}\alpha^D + C_{\bm \zeta} U C_{\bm \xi}^2 B_{\Z} B_{\x} \frac{\alpha^D - 1}{\alpha - 1}, 
			\end{align*}
			where the expectation is taken w.r.t. the distribution of $(\phi_k,y_k)$. 
		\end{lemma}
		{Note that our GNN generalization error bound is rather different from some existing results, e.g., \cite{garg2020generalization}, as edge features (i.e., $\e_{u,v}$) were not considered in their work. Lemma~\ref{lemma:generalizatio_error} can be used to understand the GNN's performance with a single batch.} 
		To characterize the node classification accuracy of the GNN learned through the described imitation learning algorithm, we need the following assumptions:
		
		\begin{assumption}\label{ass:lipschitz}
			Let $\sup_{\btheta_1, \btheta_2 \in \bm \varTheta} \| \btheta_1- \btheta_2\|_\infty \leq H$, for some $H < \infty$. Let all the loss functions $\g_i(\cdot)$ [cf. Eq.~\eqref{eq:g_def}] for $i=1,\ldots,I$ are $G$-Lipschitz continuous with respect to the $\ell_1$-norm, i.e.
			$ | \g_i(\btheta_1) - \g_i(\btheta_2)| \leq  G \|\btheta_1 - \btheta_2 \|_1, \forall i.$
		\end{assumption}

		\begin{assumption}\label{ass:realizability} The minimal empirical loss over the aggregated dataset is bounded by $\nu$.
			$$ \min_{\btheta \in \bm \varTheta} \frac{1}{IJ} \sum_{i=1}^I \sum_{(\bm \phi_s, y_s) \in \cD_i} \bbE_{\bm \psi}[\cL(\bpi_{\btheta}(\bm \phi_s), y_s)]  \leq \nu. $$
		\end{assumption}
		Assumption~\ref{ass:lipschitz} is not hard to meet if the data features and the network parameters are bounded.
		Assumption~\ref{ass:realizability} characterizes the expressiveness of the GNN. 
		
		{To present our main theory, we compute the expected number of nodes that will be visited (with the associated SOCPs/SDRs {solved}) by Algorithm \ref{algo:abb} when run with $\bpi_{\widehat{\btheta}}$ in the testing stage. Let us denote $\rho_{\widehat{\btheta}}$ as the probability with which the classifier accurately classifies a node.
			Also denote $\cS$ as the set of all possible B\&B trees that can be realized by Algorithm \ref{algo:abb} under a given instance. Let ${\rm Pr}(s; {\widehat{\btheta}}), s \in \cS$ be the probability with which a particular tree $s$ is realized. Let $Q^{s}_{\widehat{\bm \theta}}$ denote the number of visited nodes in tree $s$. Let $Q_{\widehat{\btheta}}=\mathbb{E}[Q^{s}_{\widehat{\bm \theta}}]$ where the expectation is taken over the probability mass function ${\rm Pr}(s; {\widehat{\btheta}}), s \in \cS$.
			In the following theorem, we characterize the classification accuracy, $\rho_{\widehat{\btheta}}$, and present a bound on  $\Q_{\widehat{\btheta}}$.
		}
		
		\begin{theorem}\label{thm:main}
			Suppose that Assumptions \ref{ass:lipschitz}-\ref{ass:realizability} hold, and that the GNN in \texttt{MINIMAL} is prameterized by $\widehat{\btheta}$ in \eqref{eq:theta_selection}. In addition, assume that every single batch ${\cal D}_i$ consists of i.i.d. samples, and that Algorithm \ref{algo:prune_learning} is used for GNN learning.
			Then, we have
			$$ Q_{\widehat{\btheta}} \leq  \frac{ 2N \left(2 \rho_{\widehat{\btheta}} - \rho_{\widehat{\btheta}}^N \right)}{2 \rho_{\widehat{\btheta}} - 1} + 1.$$
			Further, when $\widehat{\btheta}$ is selected using \eqref{eq:theta_selection}, with a probability at least $1- \delta$, 
			\begin{align}\label{eq:rhotheta}
				& \bbE_{p_{\widehat{\btheta}}, \bm \psi} \left[ \cL \left(\bpi_{\widehat{\btheta}}(\bm \phi_s), y_s \right) \right] \\
				& \leq \nu +  \cO \left(1/I^{1/3} \right)  + {\sf Gap}\left(\frac{\delta}{2}, J \right)\sqrt{\frac{2 \log(2/\delta)}{I}}. \nonumber
			\end{align}
			Assume the logistic loss function ${\cal L}$ is employed.
			Then, the node classification accuracy $$\rho_{\widehat{\btheta}} \geq \exp \left( -  \bbE_{p_{\widehat{\btheta}}, \bm \psi}\left[ \cL\left(\bpi_{\widehat{\btheta}}(\bm \phi_s), y_s\right)\right] \right).$$
			In addition, \texttt{MINIMAL} returns an optimal solution with probability at least $\rho_{\widehat{\btheta}}^N$.
		\end{theorem}
		The proof of Theorem \ref{thm:main} is relegated to Appendix \ref{app:proof_ml}. 
		This result bounds the number of nodes visited by the proposed algorithm under a given classification accuracy. It also characterizes the classification accuracy that can be achieved by the proposed training procedure. One can see that when the batch size is large enough, ${\sf Gap}$ is close to zero. Additionally, when the GNN is expressive (and thus $
		\nu$ is small) and the algorithm is run for large enough iterations $I$, the accuracy of the classifier, i.e., $\rho_{\widehat{\btheta}}$, approaches 1 [cf. Eq.~\eqref{eq:rhotheta}]. Consequently, the total number of nodes visited will be close to $2N+1$ at most. This shows linear dependence of the computational complexity of the proposed method on $N$, which is a significant saving compared to $N \choose L$ for the exhaustive search.
		
		\begin{remark}
			We should remark that the results in Theorem~\ref{thm:main} has a couple of caveats. First, we assumed that the samples in each ${\cal D}_i$ are i.i.d.
			If every node created by $\bm \pi_{\bm \theta^{(i)}}$ in Algorithm~\ref{algo:collect_data} is used, then the samples in ${\cal D}_i$ are likely not i.i.d., as the nodes in the same B\&B tree are generated in a sequential manner.
			Nonetheless, simple remedies can assist creating an i.i.d. batch ${\cal D}_i$---e.g., by taking only one random node from a B\&B tree. This is inevitably more costly, and seems not to be necessary in practice---as using nodes from Algorithm~\ref{algo:collect_data} for training works fairly well in our simulations. Second, the expectation based criterion \eqref{eq:theta_selection} is only approximated in practice, e.g., via using empirical averaging. Characterizing the empirical version of \eqref{eq:theta_selection} can be done via concentration theorems in a straightforward manner. However, this would substantially complicate the expressions yet reveals little to no additional insight. Hence, we leave it out of this work.
		\end{remark}

		\section{Numerical Results}\label{sec:exp}
		In this section, we showcase the effectiveness of the proposed B\&B algorithm and its machine learning based acceleration using numerical simulations.
We use CVXPY \cite{diamond2016cvxpy} which calls MOSEK \cite{aps2019mosek} to solve the SOCPs/SDRs in \eqref{eq:relaxation_node} and \eqref{eq:upper_bound}. %{Note that in all the SDR problem instances in the following experiments, rank-1 solution was obtained, hence randomization techniques were not needed.} 
{The elements of Rayleigh fading channel vectors $\{\h_m\}_{m=1}^M$ are sampled independently from circularly symmetric zero mean Gaussian distribution with unit variance. } 
Implementation of the proposed methods can be found on the authors' website\footnote{\url{https://github.com/XiaoFuLab/Antenna-Selection-and-Beamforming-with-BandB-and-ML.git}}.

\subsection{Evaluation of B\&B for Joint (R)BF\&AS}
\begin{figure}[t]
            \centering
            \includegraphics[width=0.9\linewidth]{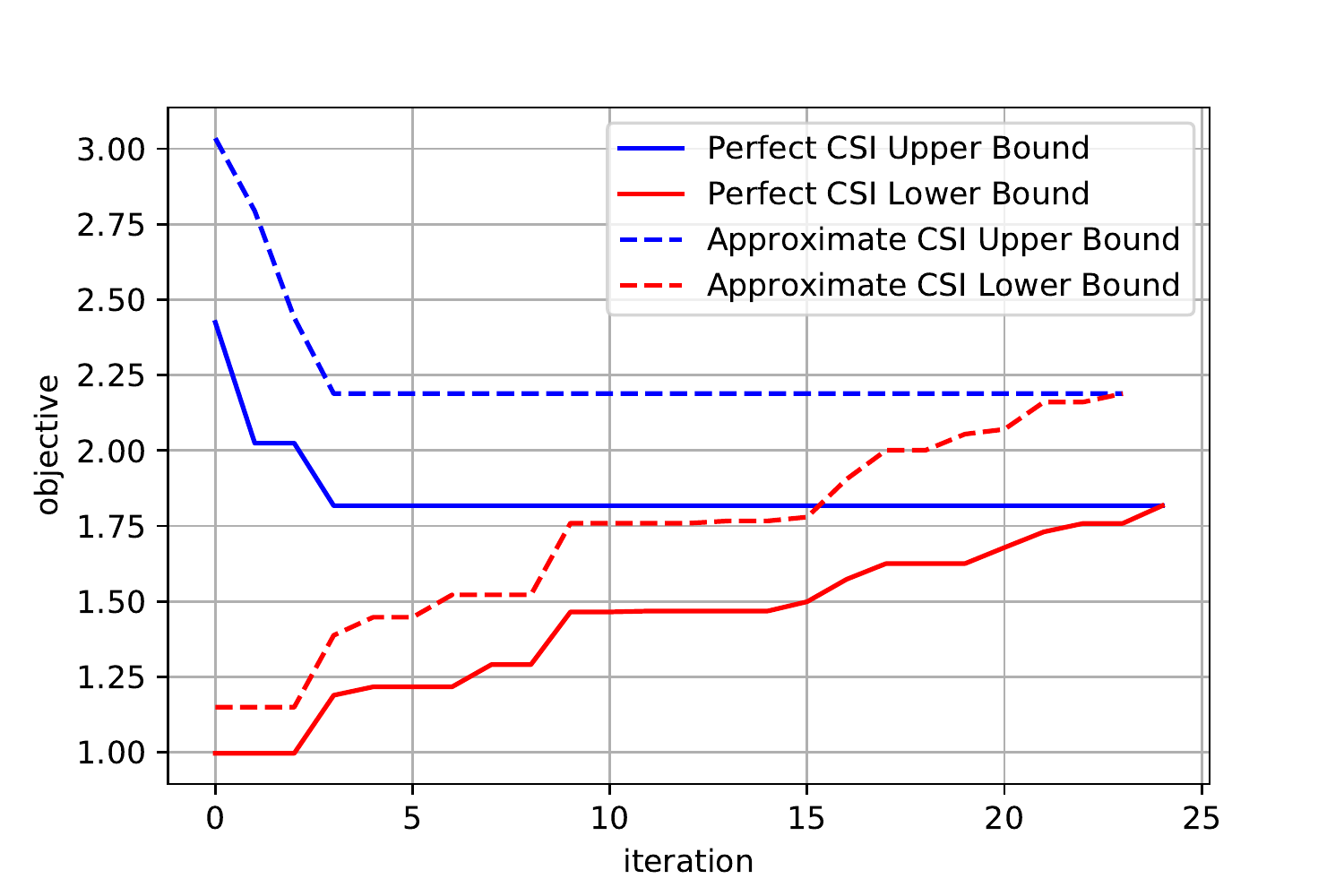}
            \caption{Convergence of the global upper and lower bounds, computed by the proposed B\&B algorithm, to the optimal solution. Problem instance of size $(N,M,L) = (8,4,4)$.}
            \label{fig:bb_convergence}
        \end{figure}

In Fig.~\ref{fig:bb_convergence}, we verify the convergence of the proposed B\&B algorithm under both the perfect and the approximate CSI cases. The figure shows the convergence of the global upper and lower bounds (i.e., $u_G^{(t)}$ and $l_G^{(t)}$) computed by the proposed B\&B procedure for $(N,M,L) = (8,4,4)$. One can see that the global bounds converge to the optimal objective value in both the perfect and approximate CSI case. 
This verifies our optimality claim in Theorem~\ref{thm:bb_convergence}.
Note that the B\&B algorithm for both cases converges in less than 24 iterations (i.e., visiting $\leq 48$ nodes). This is much less than the worst-case complexity of B\&B, i.e., visiting 139 nodes. The empirical complexity is also better than the worst-case complexity of exhaustive search, which is 70 node visits in this case.
        
Table \ref{tab:bb_performance} gives a closer look at the effectiveness of the proposed B\&B framework.
Specifically, Table \ref{tab:bb_performance} shows the performance of the proposed B\&B procedure for various problem sizes, compared to the exhaustive search strategy for the perfect CSI case. The result is averaged over 30 Monte Carlo trials. One can see that the B\&B algorithm can constantly attain reduced complexity, in terms of the number of nodes visited (i.e., the number of SOCPs solved). In particular, when the number of users is relatively small, the B\&B can attain an around 8-fold acceleration (cf. the case where $(N,M,L)=(12,2,8)$).
Similar results can be seen in Table \ref{tab:bb_approximate_performance}, where the imperfect CSI case is considered.

Table \ref{tab:alternative_form_compare} compares our B\&B and the alternative B\&B using the formulation \eqref{eq:z_formulation} in the perfect CSI case. One can see that the proposed procedure consistently solves fewer SOCPs. This supports Theorem \ref{thm:bb_convergence} (c).
        
    \begin{table}[t!]
        \caption{Performance of the proposed B\&B algorithm for various problem sizes in the perfect CSI case compared to the exhaustive search. $\sigma_m^2 = 1.0, \gamma_m=1.0, \forall m \in [M]$.}
        \centering
        \begin{tabular}{|c|c|c|c|c| }
            \hline
             Problem size  & \multicolumn{2}{|c|}{Proposed B\&B} & \multicolumn{2}{|c|}{Exhaustive Search} \\ \cline{2-5}
            $(N,M,L)$ &Time & SOCPs         & Time  & SOCPs \\ \hline 
           (8, 2, 4)   &  1.58  &   34.07   & 2.95  & 70   \\ \hline
           (8, 3, 4)   &  2.29  &   40.67   & 2.58  & 70   \\ \hline
           (8, 4, 4)   &  3.30  &   47.30   & 4.53  & 70   \\ \hline
           (8, 5, 4)   &  5.31  &   63.27   & 5.46  & 70   \\ \hline
           (8, 6, 4)   &  8.24  &   82.93   & 6.10    & 70   \\ \hline
          (10, 2, 6)   &  2.28  &    50.20  & 9.11  & 210  \\ \hline
          (10, 4, 6)   &  6.47  &   88.37   & 14.75  & 210   \\ \hline
          (10, 6, 6)   &  14.55 &    141.80 & 20.00   & 210   \\ \hline
          (10, 8, 6)   &  24.56 &    186.90 & 25.59    & 210   \\ \hline
          (12, 2, 8)   &  2.95  &    65.53  & 21.39  & 495  \\ \hline
          (12, 4, 8)   &  10.57 &   137.80  & 33.45   & 495  \\ \hline
          (12, 6, 8)   &  21.89 &    211.87 & 46.53   & 495   \\ \hline
          (12, 8, 8)   &  37.69 &    279.67 & 62.46   & 495   \\ \hline
         (12, 10, 8)   &  69.48 &    398.40 & 80.94   & 495   \\ \hline
        \end{tabular}
        \label{tab:bb_performance}
    \end{table}
    
    \begin{table}[t!]
            \caption{Performance of the proposed B\&B algorithm for various problem sizes in the Approximate CSI case compared to the exhaustive search. $\sigma_m^2 = 1.0, \gamma_m=1.0, \forall m \in [M]$.}
            \centering
            \begin{tabular}{|c|c|c|c|c| }
                \hline
                 Problem size  & \multicolumn{2}{|c|}{Proposed B\&B} & \multicolumn{2}{|c|}{Exhaustive Search} \\ \cline{2-5}
                $(N,M,L)$ &Time & SDPs    & Time &      SDPs \\ \hline 
               (8, 2, 4)  &   7.09   &    31.60 &  12.71  &  70  \\ \hline
               (8, 3, 4)  &   15.09  &    39.37 &  21.25  &  70  \\ \hline
               (8, 4, 4)  &   28.39  &    49.00 &  32.58  &  70  \\ \hline
            %   (8, 5, 4)  &   52.11  &    64.50 &    &  70  \\ \hline
              (10, 2, 6)  &   19.49  &    65.27 &  51.38  &  210  \\ \hline
              (10, 4, 6)  &   80.47  &    85.73  & 133.38   &  210 \\ \hline
              (10, 6, 6)  &   236.26 &    137.37 & 262.10   &  210 \\ \hline
              (10, 8, 6)  &   520.81 &    180.13 & 452.76   &  210 \\ \hline
              (12, 2, 8)  &   26.83  &    62.80  & 157.62   &  495 \\ \hline
              (12, 4, 8)  &   175.45 &    122.13 & 471.54   &  495 \\ \hline
            \end{tabular}
            \label{tab:bb_approximate_performance}
    \end{table}
    \begin{table}[t!]
            \centering
            \caption{Number of SOCPs solved by two B\&B Strategies. $\sigma_m^2 = 1.0, \gamma_m=1.0, \forall m \in [M]$.}
            \begin{tabular}{|c|c|c|c|c|}
            \hline
                Problem size) & \multirow{2}{*}{(4,2,2)} & \multirow{2}{*}{(8,4,6)} & \multirow{2}{*}{(8,6,6)} & \multirow{2}{*}{(10,5,6)} \\ 
                $(N,M,L)$ & & & & \\ \hline
                Proposed B\&B & 6.86 & 16.73 & 22.63 & 117.67 \\ \hline
                Alternative Using \eqref{eq:z_formulation} & 8.06 & 24.66 & 33.8 & 159.6 \\ \hline
            \end{tabular}
            \label{tab:alternative_form_compare}
        \end{table}

\subsection{Evaluation of ML-accelerated B\&B for Joint (R)BF\&AS}
In this section, we demonstrate the efficacy of \texttt{MINIMAL}. 

\subsubsection{Baselines} \label{sec:baselines}
A number of baselines are as follows:

\noindent
{\bf Supervised Learning}: We follow the supervised learning (\texttt{SL}) ideas in \cite{ibrahim2018learning,vu2021machine} to train a neural network for antenna selection (cf. Sec.~\ref{sec:existing_approaches}).  Specifically, we use the proposed B\&B algorithm to generate training pairs with optimal antenna selection as the labels, i.e., $\{\H_t, \z_t\}_{t=1}^T$, where $\z_t$ is a binary vector representing optimal antenna selection for the $t$th training instance.  The learned deep model predicts a vector $\bm z$ which may not satisfy $\|\z\|_0\leq L$, and thus we take the $L$ elements that have the largest magnitude as in \cite{ibrahim2018learning}. 
For this baseline, we use an $\f_{\btheta}$ that is a 3-layer neural network, where the first two layers are convolutional layers with ReLU activations and the last layer is a fully connected layer with sigmoid activation. 

\noindent
{\bf Greedy Method}: A plethora of greedy algorithms exist for different variants of joing BF\&AS problems; see, e.g., { \cite{chen2007efficient, mendoncca2019antenna, ding2010mmse, mahdi2021quantization,konar2018simple}.}
We design a greedy baseline for \eqref{eq:metaproblem} following the general idea of \cite{chen2007efficient}, which is described as follows: (i) Let $\cH = \{1, \dots, N\}$ denote the set of all antennas (set to active initially). (ii) Solve SOCPs with $\widetilde{\cH}_{-n}  = \cH \backslash \{n\}, \forall n \in \cH$. Let $\widehat{\cH}_{-\widehat{n}}$ correspond to the smallest objective value. Then, set $\cH = \cH \backslash \widehat{n}$. (iii) Repeat (ii) if $|\cH| > L$; otherwise return $\cH$. 

We call this method \texttt{Greedy}. Note that \texttt{Greedy}'s computational burden is not necessarily light, as a total amount of
$\cO(N^2)$ SOCPs have to be solved (e.g., $\approx 1000$ SOCPs have to be solved for $N=32$).

\noindent
{\bf Continuous Approximation}: As the third baseline, we use the continuous optimization-based idea in \cite{mehanna2013joint} and modify it to solve the unicast cases in this work. Although \cite{mehanna2013joint} did not explore their method for the approximate CSI case, we note that the same idea can be used after proper modifications to the subproblems (i.e., using the S-lemma to come up with an SDR formulation of the subproblem). We term this method \textit{iteratively reweighted convex relaxation-based optimization} (\texttt{IrCvxOpt}).

Following the implementation instruction of \cite{mehanna2013joint}, we run \texttt{IrCvxOpt} for at most 30 iterations with its bisection-based $\lambda$-tuning method for 30 iterations as well. The algorithm is stopped if the relative change of the reweighting matrix is smaller than $10^{-4}$
or a solution comprising of $\leq L$ antennas is found.  If the algorithm returns $>L$ antennas, we select the $L$ antennas from the returned antennas that is assigned the maximum power in the returned beamforming solution $\widehat{\W}$. {All of the evaluation metrics (see Sec \ref{sec:metric}) are computed using the final $L$ antennas and $\widehat{\W}$ output by the algorithms.}

\subsubsection{Training Setups}
        We use a GNN tailored for our beamforming setting 
        (see details in Appendix \ref{app:gnn}). 
        We set $(R,I)=(30,20)$ in Algorithms~\ref{algo:prune_learning}-\ref{algo:collect_data}.
       %In total 600 different problem instances are used for training. 
        %In each problem instance, $\bm H$ is generated as before.
        The loss function $\cL$ is selected to be the binary cross-entropy loss, i.e.,
        $ \cL(x,y) = -y \log(x) - (1-y) \log(1-x).$
        In batch $i$, the parameters of the classifier is initialized with $\btheta^{(i)}$, and updated using the Adam algorithm \cite{kingma2015adam} for 10 epochs, where the sample size of Adam is set to be 128. The initial step size of Adam is set to $0.001$. As described in Section~\ref{sec:imitation_learning}, we select $\widehat{\bm \theta}$ from $\btheta^{(1)}, \dots, \btheta^{(I)}$ using $30$ validation problem instances using a sample average version on \eqref{eq:theta_selection}.
        
        In order to account for the class imbalance (number of relevant nodes usually much smaller than number of irrelevant nodes in the training set), we apply a larger positive weight on the ``positive'' training pairs. Further, premature/early pruning of the B\&B tree (i.e., when $\ell$ is small) should be discouraged as it is more risky. Hence, we weight each term $\cL(\bpi_{\btheta^{(i)}} (\bm \phi_s^{(\ell)}), y_s^{(\ell)})$ using $(q \mathbb{1}[y_s^{(\ell)} = 1] + 1) \frac{1}{\ell} $,
        where $q \in \bbR$ offsets the imbalance ratio, and $\mathbb{1}[\cdot]$ denotes the indicator function. We select $q=11$ via trial and error, and use the same $q$ in all experiments.

        \subsubsection{Evaluation Metrics}\label{sec:metric}
        We define the \textit{optimality gap} (Ogap) as follows:
        $$ {\rm Ogap} := \frac{\|\widehat{\W}\|_{\rm F}^2 - \|\W^\star\|_{\rm F}^2}{\| \W^\star\|_{\rm F}^2} \times 100\%, $$
        where $\W^\star$ is the optimal solution provided by the B\&B algorithm and $\widehat{\W}$ is the solution provided by an algorithm under test. We also define the runtime speedup as follows:
        $${\rm speedup} := \frac{\text{Run-time of B\&B (seconds)}}{\text{Run-time of method under test (seconds)}}. $$
        
        \begin{table}[!t]

            \centering
            \caption{Performance of algorithms for $N\leq 16$ cases with perfect CSI. $\sigma_m^2 = 0.1, \gamma_m = 10.0, \forall m \in [M]$.}
            
            \resizebox{\linewidth}{!}{
            \begin{tabular}{|c|c|c|c|c|c|}
            % Problem & metric & proposed & baseline
            \hline
            % \rowcolor{Gray}
              Problem Size & Metric & \texttt{MINIMAL}   & \texttt{Greedy} & \texttt{IrCvxOpt} & \texttt{SL}  \\ 
            % \rowcolor{Gray}
              $(N,M,L)$ & & & & &\\ \hline
                
                \multirow{2}{*}{{${(6,3,3)}$ }}  & Ogap             & 0.00 & 1.18 & 20.54 &  64.39  \\% \cline{2-6}
                                                 & speedup          & 1.73 & 0.92 & 4.68  &  17.70    \\% \cline{2-6}
                                                 & SOCPs            & 10.25 & 15.00 & 6.65 & 1    \\ \hline
                
                \multirow{2}{*}{{${(8,4,4)}$ }}  & Ogap             & 0.0   & 0.83 & 20.19 & 38.08  \\% \cline{2-6}
                                                 & speedup          & 2.72  & 1.35 & 6.40  & 40.52 \\ %\cline{2-6}
                                                 & SOCPs            & 14.9  & 26.0 & 12.05 & 1 \\ \hline
                
                \multirow{2}{*}{{${(10,5,5)}$ }} & Ogap           & 0.85 & 2.83 & 68.34  & {-} \\ %\cline{2-6}
                                                 & speedup        & 4.10 & 2.46 & 8.47   &  {-}\\ %\cline{2-6}
                                                 & SOCPs          & 28.05 & 40.00 & 22.60 &  {-}  \\ \hline
                
                \multirow{2}{*}{{${(12,6,6)}$ }} & Ogap            & 2.16 & 3.43 & 234.88 & {-} \\ %\cline{2-6 }
                                                 & speedup         & 5.87 & 4.72 & 10.96 & {-} \\ %\cline{2-6}
                                                 & SOCPs           & 49.00 & 57.00 & 27.90 & {-}\\ \hline
                
                \multirow{2}{*}{{${(16,8,8)}$ }} & Ogap            & 2.94 & 6.59 & 159.28   & {-} \\ %\cline{2-6}
                                                 & speedup         & 12.39 & 23.88 & 78.62  & {-} \\ %\cline{2-6}
                                                 & SOCPs           & 234.50 & 100.00 & 29.00 & {-} \\ \hline
                
            \end{tabular}

            }
            \label{tab:perfect_csi_small_problem_sizes}
        \end{table}
        
        \begin{table}[!t]
            \centering
            \caption{Objective values, $\|\W\|_F^2$, attained by the algorithms for $N\geq 32$ cases with perfect CSI. $\sigma_m^2 = 0.1, \gamma_m = 10.0, \forall m \in [M]$. }
            % \caption{Comparison of the proposed method with the baselines for different problem sizes when $\sigma_m^2 = 0.1$ and $\gamma_m = 10.0$. For $N\geq 32$, we limit the number of SOCPs that is allowed to be solved by all methods to $2N$. If the solutions returned by the methods do not satisfy $\|\widehat{\W}\|_{\rm row-0} \leq L$, we pick the $L$ antennas that are assigned maximum power in $\widehat{\W}$. }
%             {'ogap': [12.441837742744763,
%   72.7354171252774,
%   3.1363700353904496,
%   163.93844947551347],
%  'feasibility': [1.0, 1.0, 1.0, 1.0],
%  'speedup': [],
%  'num_problems': [64.0, 128.0, 256.0, 256.0],
%  'time': []}
            \resizebox{0.8\linewidth}{!}{
            \begin{tabular}{|c|c|c|c|}
            % Problem & metric & proposed & baseline
            \hline
            % \rowcolor{Gray}
              Problem Size  & { \texttt{MINIMAL}}   & \texttt{Greedy} & \texttt{IrCvxOpt}   \\ 
            % \rowcolor{Gray}
              $(N,M,L)$ & & & \\ \hline
                
                % \multirow{2}{*}{{${(32,12,12)}$ }} & $\|\W\|_F^2$   & 4.35 & 21.73 & 12.62  \\ %\cline{2-5}
                %                                  & SOCPs            & 66.00 & 66.00 & 30.00    \\ \hline
                % % \rowcolor{Gray}
                % % \multicolumn{5}{|c|}{} \\ \hline
                
                % \multirow{2}{*}{{${(64,16,16)}$ }} & $\|\W\|_F^2$   & 5.23 & 61.66 & 10.72   \\ %\cline{2-5}
                %                                  & SOCPs            & 130.00 & 130.00 & 30.00    \\ \hline
                % % \rowcolor{Gray}
                % % \multicolumn{5}{|c|}{} \\ \hline
                
                % \multirow{2}{*}{{${(128,8,8)}$ }} & $\|\W\|_F^2$   & 1.86 & 22.45 & 3.14  \\ %\cline{2-5}
                %                                  & SOCPs            & 258.00 & 258.00 & 30.00      \\ \hline
        
                % \multirow{2}{*}{{${(128,16,16)}$ }} & $\|\W\|_F^2$   & 4.60 & 40.29 & 166.69  \\ %\cline{2-5}
                %                                  & SOCPs            & 258.00 & 258.00 & 30.00     \\ \hline

                \multirow{1}{*}{{${(32,12,12)}$ }}   & 4.35 & 21.73 & 12.44  \\ %\cline{2-5}
                                                %  & SOCPs            & 66.00 & 66.00 & 30.00    \\ \hline
                % \rowcolor{Gray}
                % \multicolumn{5}{|c|}{} \\ \hline
                
                \multirow{1}{*}{{${(64,16,16)}$ }}   & 5.23 & 61.66 & 72.73   \\ %\cline{2-5}
                                                %  & SOCPs            & 130.00 & 130.00 & 30.00    \\ \hline
                % \rowcolor{Gray}
                % \multicolumn{5}{|c|}{} \\ \hline
                
                \multirow{1}{*}{{${(128,8,8)}$ }}  & 1.86 & 22.45 & 3.13  \\ %\cline{2-5}
                                                %  & SOCPs            & 258.00 & 258.00 & 30.00      \\ \hline
        
                \multirow{1}{*}{{${(128,16,16)}$ }}   & 4.60 & 40.29 & 163.93  \\ %\cline{2-5} 
                                                %  & SOCPs            & 258.00 & 258.00 & 30.00     \\ \hline
        \hline
            \end{tabular}
            }
            \label{tab:perfect_csi_large_prob_sizes}
        \end{table}
        
        \subsubsection{Results}

        { 
        Table \ref{tab:perfect_csi_small_problem_sizes} shows the performance of all methods under $\gamma_m = 10.0, \sigma_m^2 = 0.1, \forall m \in [M]$ for cases where $N\leq 16$. Results are averaged over 20 random test instances. 
        One can see that { \texttt{MINIMAL}} consistently attains a very small Ogap ($<3\%$ for all cases), whereas the baselines have much larger Ogaps. The \texttt{SL} method only requires solving a single SOCP, as the antenna selection part is done by the learned $\f_{\widehat{\btheta}}$. However, the solution quality is not acceptable, indicating that the learned neural network for AS performs poorly. Notably, in our simulations, we observed that \texttt{SL} needs a large amount of problem instances to generate its training data for a given $(N,M,L)$.
        For example, under the settings in Table~\ref{tab:perfect_csi_small_problem_sizes}, $T=12,000$ instances were used for \texttt{SL}, but only $600$ instances were used for the proposed method.

        Table \ref{tab:perfect_csi_large_prob_sizes} shows the performance of the algorithms in cases where $N \geq 32$. 
        Note that generating training samples for \texttt{SL} is too costly in these case, and thus we drop this baseline in this table. This is because for each $(N,M,L)$, one has to re-train $\f_{\btheta}$ from scratch under \texttt{SL}---but generating training examples for large size $N$ is not affordable. 
        For the proposed algorithm, we use the GNN trained on smaller problem size, i.e., $(N,M,L) = (16,8,8)$ (cf. Remark~\ref{rmk:gnn}), which allows us to avoid re-training.
        In this simulation, we test all methods under limited computational budget (i.e., every method is allowed to use up to $2N$ SOCPs), for controlling the runtime.
        Unlike the previous cases where the Ogap is presented, we could only compare the objective values in this simulation, as obtaining the optimal solution is very costly. 
        One can see that the proposed method attains objective values that are oftentimes order-of-magnitude smaller than those of the baselines. \texttt{IrCvxOpt} sometimes attains small objective values (e.g., when $(N,M,L)=(128,8,8)$), but the performance is not consistent across different cases.
        }

        \begin{table}[!t]
         
        \centering
        \caption{
        Performance of algorithms under approximate CSI. $\sigma_m^2 = 0.1, \gamma_m=10.0, \varepsilon_m= 0.02, \forall m \in [M]$. }
        \resizebox{\linewidth}{!}{
        \begin{tabular}{|c|c|c|c|c|c|}
        \hline
          Problem Size & Metric & { \texttt{MINIMAL}}  & \texttt{Greedy} & \texttt{IrCvxOpt} & \texttt{SL}   \\ 
          $(N,M,L)$ & & & & &\\ \hline

            \multirow{2}{*}{{${(8,4,4)}$ }} & Ogap & 0.09 & 1.27 & 4.97 & 21.97  \\ %\cline{2-6}
                                          & speedup & 3.54 & 1.43 & 10.64 & 47.08 \\ %\cline{2-6}
                                          & SDRs & 13.30 & 26.00 & 4.70 & 1.0    \\ \hline

            \multirow{2}{*}{{${(10,5,5)}$}}   & Ogap & 2.04 & 2.20 & 10.72    & {-}\\ %\cline{2-6}
                                              & speedup & 4.19 & 1.90 & 18.89 & {-}\\ %\cline{2-6}
                                          & SDRs & 23.90 & 40.00 & 7.75       & {-}\\ \hline
            
            \multirow{1}{*}{{${(16,8,8)}$}}   & $\|\W\|_F^2$ & 2.93 & 24.39 & 3.15 & {-}\\ %\cline{2-6}
                                          & SDRs & 34.00 & 34.00 & 18.25      & {-}\\ \hline
            
        \end{tabular}
        \label{tab:approximate_csi_prob_sizes}
        }
        
    \end{table}

        { Table \ref{tab:approximate_csi_prob_sizes} shows the performance of the algorithms under imperfect CSI using the RBF constraints. %{ Result is averaged over 20 random trials.
        For $(N,M,L) = (16,8,8)$, we use the model trained on $(N,M,L) = (10,5,5)$, and limit the number of SDRs to $2N$.
        Similar to the perfect CSI case, the proposed method attains the smallest Ogap/objective value compared to all baselines. 
        The \texttt{IrCvxOpt} again sometimes outputs acceptable results, but could not maintain consistently good performance over all cases. 
        
        }
    
      \begin{table}[t]
        \centering
        \caption{Performance of Algorithms under Various $\gamma_m$'s  with Approximate CSI. $(N,M,L) = (8,4,4)$, $\varepsilon_m = 0.02, \sigma_m^2 = 0.1, \forall m \in [M]$. } 
        \resizebox{\linewidth}{!}{
        \begin{tabular}{|c|c|c|c|c|}
             % In dB 30, 33.01, 34.77, 36.02, 36.98, 37.40
             % bits/s/Hz 9.96, 10.96, 11.55, 11.96
            \hline
              $\gamma_m$(dB) & Metric   & { \texttt{MINIMAL}}  & \texttt{Greedy} & \texttt{IrCvxOpt}   \\ 
              (\# feasible ins.) & & & & \\ \hline 
                    %9.96 bits/s/Hz 
                    30.00 & Ogap  & 0.40 & 4.63 & 17.76    \\ %\cline{2-6}
                    (50) & \# feasible solutions & $50$    & $50$    & $44$   \\   \hline%\cline{2-6}
                
                    % 10.96 bits/s/Hz     
                    33.01  & Ogap  & 0.51 & 11.21 & 45.07     \\ %\cline{2-6}
                    (40) & \# feasible solutions & $40$    & $39$   & $32$   \\    \hline   %\cline{2-6}
                
                    % 11.55 bits/s/Hz     
                    34.77  & Ogap & 0.00 & 19.02 & 133.88     \\ %\cline{2-6}
                    (25) & \# feasibile solutions   & $25$    & $25$   & $21$   \\ \hline
                                                
                    % 11.96 bits/s/Hz     
                    36.02  & Ogap & 0.00 & 72.19 & 31.65    \\ %\cline{2-6}
                    (10) & \# feasible solutions &   $10$   & $10$   & $7$   \\ \hline
                    
                    % \multirow{2}{*}{5000}  {\purple 12.28 bits/s/Hz}     & Ogap   & 0.00 & 0.00 & 0.00     \\ %\cline{2-6}
                    %                             & $p/f/t$                    & $3/3/50$ & $2/3/50$   & $2/3/50$   \\ \hline
                                                
                    % \multirow{2}{*}{5000}  {\purple 12.28 bits/s/Hz}     & Ogap   & 0.00 &  5.84 & 0.00     \\ %\cline{2-6}
                    %                             & $p/f/t$                    & $8/8/100$ & $8/8/100$   & $7/8/100$   \\ \hline

            \end{tabular}
            }
        \label{tab:feasibility}
    \end{table}

{ 
        Table \ref{tab:feasibility} tests the algorithms' ability of finding feasible solutions of \eqref{eq:metaproblem}.
        Note that finding a feasible solution for QCQP problems is often highly nontrivial \cite{mehanna2014feasible}.
        As making $\|\W\|_{\rm row-0}\leq L$ [cf. Eq.~\eqref{eq:cardinality}] can be easily done via simple post-processing (e.g., by thresholding some rows of the solution $\W$ to zeros), we primarily examine if the algorithms could find $\W$'s that satisfy the SINR specifications in \eqref{eq:sinr_constraint}.        
        To be specific, the algorithms are tested using various $\gamma_m$'s. Naturally, higher values of $\gamma_m$ may make all the SINR constraints hard to satisfy. We run 50 random trials. 
        One can see that under {$\gamma_m=$30dB}, all the problem instances have at least a feasible solution for \eqref{eq:sinr_constraint}.
        Both { \texttt{MINIMAL}} and \texttt{Greedy} can find solutions that are feasible for all instances, but { \texttt{MINIMAL}} enjoys a much smaller Ogap.
        When $\gamma_m$ grows, the problem admits fewer infeasible instances. However, { \texttt{MINIMAL}} always returns a feasible solution, as long as the instance has one. \texttt{Greedy} also works fine for finding feasible solutions, but the Ogap becomes much larger when $\gamma_m$ increases. \texttt{IrCvxOpt} is less competitive in terms of both Ogap and feasibility. 
    
     \section{Conclusion and Discussion}}
         In this work, we revisited the joint beamforming and antenna selection problem under perfect and imperfect CSI and proposed a machine learning-assisted B\&B algorithm to attain its optimal solution. Unlike the vast majority of existing algorithms that rely on continuous optimization to approximate the hard mixed integer and nonconvex optimization problem without optimality guarantees, our B\&B algorithm leverages the special properties of joint (R)BF\&AS to come up with optimal solutions. More importantly, we proposed a GNN-based machine learning method to help accelerate the B\&B algorithm. Our analysis showed that the design ensures provable acceleration and retains optimality with high probability, under proper GNN design and given a sufficiently enough sample size. To our best knowledge, this is the first comprehensive characterization for ML-based B\&B. Our GNN design also easily handles a commonly seen challenge in communications, namely, the problem size change across training and test sets, without visible performance losses. Simulations corroborated our design goals and theoretical analyses.

     {Moving forward, a natural question is if the proposed ML-accelerated B\&B method can be extended to offer efficient and optimal solutions to other joint (R)BF\&AS criteria, e.g., those in \cite{golbon2016beamforming,dai2006optimal,wang2014outage,mehanna2013joint,shi2014group,ibrahim2020fast}.  
     This can {\it in principle} be done, but the caveat lies in designing an effective B\&B algorithm for the problem of interest. In our case, our B\&B design leveraged the fact that \eqref{eq:metaproblem} is optimally solvable when given a fixed set of antennas, which is a property that not all the joint BF\&AS formulations enjoy---e.g., the multicast version of \eqref{eq:metaproblem} cannot be handled by a similar B\&B. Therefore, a meaningful future direction is to consider such more challenging cases and come up with a ML-assisted (near)-optimal method.
     }
     
    %  Finally, a limitation of the B\&B  is designed for QoS formulation. Since the B\&B design has many problem specific components, a different problem formulation, e.g., sum-rate maximization \cite{dai2006optimal}, requires a separate B\&B design. 
     
    %  will be necessary since  while the GNN-design can be  the proposed B\&B design is specific to \eqref{eq:metaproblem}. There is no unified recipe that works for all problems. A different problem formulation, e.g., sum-rate maximization \cite{dai2006optimal}, may require a separate B\&B design.  

    % {\red Sagar: do not highlight the newly cited papers ...}
    %  {\orange It is not highlighted}


		\appendices
		
		\section{Proposed B\&B Procedure}\label{app:bb_procedure}
		The proposed B\&B procedure is essentially Algorithm~\ref{algo:abb} without any pruning of the nodes based on node classifier. The B\&B procedure is outlined in Algortihm~\ref{algo:bb}
		
		\begin{algorithm}[h]\label{algo:bb}
			\footnotesize
			\SetAlgoLined
			Input: Problem instance $(\h_m, \sigma_m, \gamma_m, \varepsilon_m), \forall m$, trained pruning policy $\bpi_{\btheta}$, relative error $\epsilon$; \\
			\tcp{Add the root node first}
			$\cA^{(0)}_1 \leftarrow \{\}, \cB^{(0)}_1 \leftarrow \{\} $; \\
			Select node using \eqref{eq:node_selection} for $\cN^{(0)}_1$; \\
			$\W_{\rm incumbent} \leftarrow$ solution to \eqref{eq:upper_bound}; \\
			$l_G^{(t)} \leftarrow \|\W^{(0)}_1\|_F^2$, $u_G^{(0)} \leftarrow \|\W_{\rm incumbent}\|_F^2$; \\
			$\cF^{(0)} \leftarrow \{ (0,1) \}$; \\
			$t \leftarrow 0$; \\
			\While{$|\cF^{(t)}|> 0$ and $\nicefrac{\left|u_G^{(t)} - l_G^{(t)}\right|}{l_G^{(t)}} > \epsilon$ 
			}{
				Select a non-leaf node $(\ell^\star, s^\star)$ using \eqref{eq:node_selection} \\
				Remove the selected node $ \cF^{(t)} \leftarrow \cF^{(t)} \backslash \cN^{(\ell^\star)}_{s^\star} $; \\ 
				
				Select variable $n^\star$ using \eqref{eq:variable_selection};\\
				% Generate child nodes 
				Generate child nodes $\cN^{(\ell^\star+1)}_{s^\star_1}$ and $\cN^{(\ell^\star+1)}_{s^\star_2}$ using \eqref{eq:child_node_generation} and append to $\cF^{(t)}$; \\
				$k \leftarrow \arg \min_{i \in \{ 1, 2\}} \Phi_{\rm ub}\left(\cN^{(\ell^\star + 1)}_{s^\star_i}\right)$; \\
				\If{$\Phi_{\rm ub}\left(\cN^{(\ell^\star + 1)}_{s^\star_k}\right) \leq u_G^{(t)}$}
				{
					$u_G^{(t+1)} \leftarrow  \Phi_{\rm ub}\left(\cN^{(\ell^\star + 1)}_{s^\star_k}\right)$; \\
					$\W_{\rm incumbent} \leftarrow $ solution to \eqref{eq:upper_bound} for $\cN^{(\ell^\star + 1)}_{s^\star_k}$; \\
				}
				$l_G^{(t+1)} \leftarrow  {\rm min}_{(\ell, s) \in \cF^{(t)}} \Phi_{\rm lb}\left(\cN^{(\ell)}_s\right)$; \\
				
				${\cal F}^{(t+1)} \leftarrow \left\{(s',\ell') \in {\cal F}^{(t)}~|~ \Phi_{\rm lb}\left({\cal N}_{s'}^{(\ell')}\right) \leq u_G^{(t+1)} \right\};$ \\
				
				$t \leftarrow t + 1$; \\
			}
			Return $\W_{\rm incumbent}$; \\
			\caption{\texttt{BB} }
		\end{algorithm}

		\section{Poof of Lemma \ref{lem:solvinglowerbound}}\label{app:proof_lemmas}
		\textbf{(a)} The BF setting implies that $\cC(\w_m, \h_m, \varepsilon_m, \sigma_m)$ is from \eqref{eq:sinr_bf}. Then, the equivalence of \eqref{eq:sinr_bf} and \eqref{eq:bf_convex_reformulation} implies that \eqref{eq:relaxation_node} for any node $\cN_s^{(\ell)}$ can be optimally solved using SOCP. Hence Lemma \ref{lem:solvinglowerbound}(a) holds due to Lemma~\ref{lemma:socp}. 
		
		\textbf{(b)} Note that \eqref{eq:relaxation_node} with $\cB_s^{(\ell)}$ is equivalent to \eqref{eq:robust_beamforming} with antennas restricted to the set $[N]\backslash \cB_s^{(\ell)}$. Hence, when the condition in \eqref{eq:cond_rbf_subset} is satisfied for $\H([N] \backslash {\cB_s^{(\ell)}}, :)$, then \eqref{eq:relaxation_node} with $\cB_s^{(\ell)}$ can be optimally solved using SDR due to Lemma \ref{lemma:optimal_sdr}. Further, the B\&B procedure ensures that $|\cB_s^{(\ell)}| \leq N-L, \forall (s, \ell)$. Hence, the set $\{\H({\cS},:) | \cS \in [N], |\cS| \geq L\}$ includes all possible instances of \eqref{eq:relaxation_node} encountered during the B\&B procedure. Therefore, Lemma~\ref{lem:solvinglowerbound}(b) holds.
		
		{
			\textbf{(c)} Note that $|\widetilde{\cB}_s^{(\ell)}| = N-L$. Hence, the solution of Problem \eqref{eq:upper_bound} satisfies the constraint \eqref{eq:cardinality}. Further, due to Lemma \ref{lem:solvinglowerbound} (a) and (b), Problem \eqref{eq:upper_bound} can be optimally solved using SOCP and SDR for the BF and RBF cases, respectively. Hence, $\Phi_{\rm ub}(\cN_s^{(\ell)})$ is a valid upper bound of the optimum of \eqref{eq:metaproblem}.
		}

		\section{Proof of Theorem \ref{thm:bb_convergence}}\label{app:proof_bb}
		% Let $(\cA^\star, \cB^\star)$ be the optimal antenna decision.
% In each iteration, one of the following two cases occur:

\subsection{Proof of (a) and (b)}
{
Note that if the SOCP and SDR return optimal solutions to every leaf node of the B\&B tree, then the B\&B procedure is ensured to find the optimal solutions of the the joint BF/RBF\&AS problems. The reason is that the B\&B tree only has a finite number of leaves. %Hence, there exists a leaf that must return the global optimal solution, since the leaves represent the situations where exactly $L$ antennas are activated.

For the BF setting with perfect CSI, the subproblem at a leaf node $(\ell,s)$ can be expressed as 
\begin{align}  \label{eq:subnodesocp}
        \minimize_{\W} ~ & \|\W\|_F^2  \\
     \text{subject to } ~ &  \frac{|\w_m^H\h_m|^2}{\sum_{\ell \not = m} |\w_\ell^H \h_m|^2 + \sigma_m^2 } \geq \gamma_m, \quad \forall m \in [M] \nonumber \\
        & \W(n, : ) = \zero, \quad\forall n \in {\cal B}^{(\ell)}_s, \nonumber 
\end{align}
where $|{\cal B}^{(\ell)}_s| = N-L$. Since $\|\W\|_{{\rm row}-0} \leq L$ is automatically satisfied, it is omitted. Problem~\eqref{eq:subnodesocp} can be rewritten as
\begin{align}  \label{eq:subnodesocp_mod}
        \minimize_{\W_s^{(\ell)}} ~ & \|\W_s^{(\ell)}\|_F^2  \\
     \text{subject to } ~ &  \frac{|\w_m^H\h_m|^2}{\sum_{\ell \not = m} |\w_\ell^H \h_m|^2 + \sigma_m^2 } \geq \gamma_m, \quad \forall m \in [M] \nonumber 
\end{align}
where $\W_s^{(\ell)}=\W([N] \backslash {\cal B}_s^{(\ell)},:)$, and we let $\w_m = \W_s^{(\ell)}(:,m)$ by slightly abusing the notation.
Since Problem~\eqref{eq:subnodesocp_mod} can be recast as a convex problem as detailed in \eqref{eq:bf_convex_reformulation}, the solution to the above is indeed optimal.

Similarly, under the RBF setting with imperfect CSI, the subproblem at a leaf node can be written as
\begin{align}\label{eq:subnodesdr}
        \minimize_{\W} ~ & \|\W\|_F^2  \\
     \text{subject to } ~ &  \min_{\overline{\h}_m \in \cU_m} \frac{\overline{\h}_m^H \W_m \overline{\h}_m}{\sum_{j \not= m} \overline{\h}_m^H \W_j \overline{\h}_m + \sigma_m^2} \geq \gamma_m,  \nonumber \\
        & \W(n, : ) = \zero, \quad\forall n \in {\cal B}^{(\ell)}_s, \nonumber 
\end{align}
where $|{\cal B}^{(\ell)}_s| = N-L$. Problem \eqref{eq:subnodesdr} can be further rewritten as 
\begin{align}\label{eq:subnodesdr_mod}
    \minimize_{\W_s^{(\ell)}} ~ & \|\W_s^{(\ell)}\|_F^2  \\
     \text{subject to } ~ &  \min_{\overline{\h}_m \in \cU_m} \frac{\overline{\h}_m^H \W_m \overline{\h}_m}{\sum_{j \not= m} \overline{\h}_m^H \W_j \overline{\h}_m + \sigma_m^2} \geq \gamma_m, \nonumber
\end{align}
where $\W_s^{(\ell)}$ and $\w_m$ are defined as in \eqref{eq:subnodesocp_mod}, and $\h_m = \H_s^{(\ell)}(:,m)$ with $\H_s^{(\ell)}= \H ([N] \backslash \cB_s^{(\ell)}, :)$ (recall that $\cU_m := \{\h_m + \e | \|\e\|_2 \leq \varepsilon_m \} $). Using the condition in Theorem~\ref{thm:bb_convergence} (b), and invoking Lemma~\ref{lem:solvinglowerbound}, one can see that \eqref{eq:subnodesdr_mod} can be solved optimally using SDR.
}

\subsection{Proof of (c)}
\subsubsection{Amount of SOCPs/SDRs Solved by Proposed B\&B}
In our B\&B procedure, \eqref{eq:relaxation_node} and \eqref{eq:upper_bound} are equivalent for any node and its right child node, i.e., 
\begin{align*}
 \Phi_{\rm lb}\left(\cN_s^{(\ell)}\right) = \Phi_{\rm lb}\left(\cN_{s_2}^{(\ell + 1)}\right), \Phi_{\rm ub}\left(\cN_s^{(\ell)}\right) = \Phi_{\rm ub}\left(\cN_{s_2}^{(\ell + 1)}\right).  
\end{align*}
The first equation is because $\cB_{s}^{(\ell)} = \cB_{s_2}^{(\ell +1)}$ and the second because $\widetilde{\cB}_s^{(\ell)}, \forall (\ell, s)$ in \eqref{eq:upper_bound} is determined using the solution to \eqref{eq:relaxation_node}. 
Hence, one can avoid redundant computations in the nodes by storing and reusing the results of \eqref{eq:relaxation_node} and \eqref{eq:upper_bound}. 
Using this observation, we derive an upper bound of the number of SOCPs/SDRs that need to be solved by the B\&B. 

\begin{figure}
    \centering
    \includegraphics{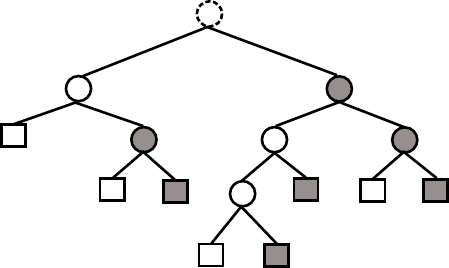}
    \caption{Illustration of a B\&B tree (where no nodes are fathomed). }
    \label{fig:counting_redundant}
\end{figure}

Consider a B\&B tree where none of the nodes are fathomed (Fig. \ref{fig:counting_redundant}). Note that there are $Q_{\rm Leaf} = {N \choose L}$ leaf nodes (squares in Fig. \ref{fig:counting_redundant}). 
Therefore, there are $Q_{\rm Total} = 2{N \choose L} -1$ nodes in total (all circles and squares). 
% Hence, there are ${N \choose L} - 1$ non-leaf nodes (all circles). 
Each non-leaf node (circles) is branched into a right child node and a left child node. Hence, there are  $Q_{\rm Right} = {N \choose L} - 1$ right child nodes (shaded solid circles and squares) and $Q_{\rm Left} = {N \choose L} - 1$ left child nodes (unshaded solid circles and squares). 

The constraints of the SOCPs/SDRs corresponding to the leaf nodes can be different from that of its parent even if they correspond to a right child node, i.e., shaded squares. 
This is because of the update step in \eqref{eq:determine_node} for the leaf nodes. To explain, a right child node, $\cN_s^{(\ell)}$, is converted into a leaf node if $L$ of the decided antennas are included, i.e., $|\cA_s^{(\ell)}| = L$. For this node, $\cB_s^{(\ell)} = [N] \backslash \cA_s^{(\ell)},$
i.e., all remaining undecided antennas are excluded. Since $\cB_s^{(\ell)}$ will be different from that of its parent node, the solutions of \eqref{eq:relaxation_node} and \eqref{eq:upper_bound} can be different from that of its parent node.

Therefore, only the non-leaf right child nodes (shaded solid circles) can reuse previously stored upper bound and lower bound solutions from their parents. Let $Q_{\rm RightLeaf}$ denote the number of right child leaf nodes (shaded squares). Then, the total number of nodes whose associated SOCPs/SDRs that need to be solved in the worst case is
$ Q_{\rm Compute} =  Q_{\rm Total} - Q_{\rm Right} + Q_{\rm RightLeaf}.$

To count $Q_{\rm RightLeaf}$, notice that the right and left child nodes of a parent node correspond to `including' and `excluding' an antenna, respectively. A parent node is branched into a right child leaf node if it contains exactly $L-1$ included antennas and fewer than or equal to $ N-L-1$ excluded antennas. This implies that there can be fewer than or equal to $(L-1) + (N-L-1) = N-2$ decided antennas. Hence, a right child leaf node is created whenever a node has  $\leq N-2$ decided antennas, where $L-1$ of them are included, is branched. Therefore, we have the following holds:
\begin{align*}
     Q_{\rm RightLeaf} &= {N-2 \choose L-1} + {N-3 \choose L-1} + \dots + {L-1 \choose L-1} \\
    &= \sum_{i=2}^{N-L+1} {N-i \choose L-1}.
\end{align*}

Consequently, 
$$ Q_{\rm Compute} = {N \choose L} + \sum_{i=2}^{N-L+1} {N-i \choose L-1}.$$
Note that $Q_{\rm Compute}$ nodes may correspond to $2Q_{\rm Compute}$ SOCPs/SDRs (cf. \eqref{eq:relaxation_node} and \eqref{eq:upper_bound} for each node). However, for the leaf nodes \eqref{eq:relaxation_node} and \eqref{eq:upper_bound} are identical. Hence there are only $Q_{\rm Compute} - {N \choose L}$ instances of \eqref{eq:relaxation_node}. Moreover, there can be at most $N \choose L$ instances of \eqref{eq:upper_bound}, since ${N \choose L}$ correspond to selecting $L$ out of $N$ antennas. Therefore, there are at most $Q_{\rm Compute}$ SDRs/SOCPs solved by the B\&B procedure.

\subsubsection{The SOCPS/SDRs Needed in B\&B for Problem \eqref{eq:z_formulation}}\label{app:alternative_formulation}
To complete the proof, let us examine the number of SOCPs/SDRs that are needed to exhaust the B\&B tree of the formulation in \eqref{eq:z_formulation}.

A node problem of \eqref{eq:z_formulation}, for the node $\cN_s^{(\ell)}$ is as follows:
\begin{align}\label{eq:minp_node_problem}
    \minimize_{\W, \z} &~\|\bm W\|_F^2 \\
    {\rm subject~to}&~{\cal C}(\bm w_m,\bm h_m, \varepsilon_m, \sigma_m) \geq \gamma_m,  \nonumber \\
    &~ \z \in \{0, 1\}^N, ~ \z^\T \one \leq L, \nonumber \\
    &~ z(n) = 0, ~ n \in \cB_s^{(\ell)},~ z(n) = 1, ~ n \in \cA_s^{(\ell)}, \nonumber \\
    &~ \|\W(n,:)\|_2 \leq C z(n), ~\forall n \in [N]. \nonumber
\end{align}
The lower bound is obtained by solving the convex relaxation of the above, i.e., $\z \in \{0,1\}$ is relaxed to $\z \in [0,1]^N$. 
One can see that the lower bounds obtained at the parent node and both child nodes may be different.

It is because \eqref{eq:minp_node_problem} depends upon both $\cA_s^{(\ell)}$ and $\cB_s^{(\ell)}$ and each child node will differ from its parent in one of the two sets, i.e, $\cB_{s_1}^{(\ell+1)} \neq \cB_{s}^{(\ell)}$ and $\cA_{s_2}^{(\ell+1)} \neq \cA_{s}^{(\ell)}$. 
The above implies that the number of SOCPs/SDRs with B\&B using \eqref{eq:minp_node_problem} has an upper bound of $ Q'_{\rm Compute} = 2{N \choose L} - 1$ (specially, with ${N \choose L}$ instances of \eqref{eq:upper_bound} and ${N \choose L} - 1$ instances of \eqref{eq:relaxation_node}).

\vfill
\begin{mdframed}
Due to page limitations,
the proofs of Lemmas 4-6 and Theorem 2, the details of the GNN design, and the details of node feature design
can be found in the supplementary materials (download under ``Media'' of the IEEE Xplore page of this paper). They can also be found online at \url{https://arxiv.org/abs/2206.05576}.

\end{mdframed}
		
		\clearpage
		
		{\bf Supplementary Material of ``Optimal Solutions for Joint Beamforming and Antenna Selection: From Branch and Bound to Machine Learning'' \\
			\begin{center}
				Sagar Shrestha, Xiao Fu, and Mingyi Hong
			\end{center}
		}
		
		\section{Proof of Lemma \ref{lemma:generalizatio_error}}

We use the empirical Rademacher complexity of the GNN class to assist finding the expected risk's error, which is a classic way of establishing generalization bounds \cite{mohri2018foundations, bartlett2017spectrally, chen2019generalization}. To proceed, let us define the sets
\begin{align*}
\cX_{\bm \phi} &:= \big\{ \bm \phi = \left[\x_1^\T, \dots, \x_U^\T, \e_{1,1}^\T, \dots, \e_{U,U}^\T \right] \\
& ~\big|~ \|\x_u\|_2, \|\e_{u,v}\|_2 \leq B_{\x}, \forall u,v \in [U] \big\}, \\
\cX_{\Z} &:= \{ \Z \in \bbR^{E \times E}| \|\Z\|_2 \leq \cB_{\Z}\}, \text{ and}\\
\cX_{\bm \beta} &:= \{ \a \in \bbR^E ~|~ \|\a\|_2 \leq B_{\bm \beta}\}.
\end{align*}
First, consider the following lemma:
\begin{lemma}[{\cite[Theorem 3.1]{mohri2018foundations}}]\label{lemma:rademacher} 
Let $\cT$ be a family of functions mapping from $\cX_{\bm \phi} \times \{0,1\}$ to $[-b,b]$. 
Assume ${\cal G}$ consists of $K$ i.i.d. samples $\{\bm \phi_k,y_k\}_{k=1}^K$.
With probability at least $1-\delta$ over the samples $\cG$, for any $\tau \in \cT$, 
\begin{align*}
 & \bbE[\tau(\bm \phi, y)] - \frac{1}{K} \sum_{(\bm \phi_k, y_k) \in \cG} \tau(\bm \phi_k, y_k)  \leq 2 {\widehat{\cR}_{\cG}(\cT)} + 3 b\sqrt{\frac{\log{2/\delta}}{2K}},
\end{align*}
where ${\widehat{\cR}_{\cG}(\cT)}$ is the empirical Rademacher complexity \cite{mohri2018foundations} of the set $\cT$ with respect to the samples $\cG$. 
\end{lemma}
Let us define the set $\cT := \{(\bm \phi, y) \mapsto \cL(\bpi_{\btheta}(\bm \phi), y) ~|~ \btheta \in \bm \varTheta \}$, a class of functions that maps from $\cX_{\bm \phi} \times \{0,1\}$ to $[-B_{\cL}, B_{\cL}]$. Then, applying Lemma \ref{lemma:rademacher} to $\cT$ over the set $\cG$ ensures that with probability at least $1-\delta$ over $\cG$, $\forall \btheta \in \bm \varTheta$,    
\begin{align}\label{eq:generalization_error}
 & \bbE[\cL(\bpi_{\btheta}(\bm \phi), y)] - \frac{1}{K} \sum_{i \in [K]} \cL(\bpi_{\btheta}(\bm \phi_i), y_i)  \nonumber \\
 & \leq 2 {\widehat{\cR}_{\cG}(\cT)} + 3 B_{\cL}\sqrt{\frac{\log{2/\delta}}{2K}},
\end{align}
In the following, we derive an upper bound on $\widehat{\cR}_{\cG}(\cT)$. To this end, we instead define a set $\bm \Pi := \{ \bm \phi \mapsto \bpi_{\btheta}(\bm \phi) | \btheta \in \bm \varTheta \}$, and derive $\widehat{\cR}_{\cG}(\bm \Pi)$. With this, we can use Talagrand's Lemma \cite[Lemma 4.2]{mohri2018foundations} to obtain $\widehat{\cR}_{\cG}(\cT)$ as $\widehat{\cR}_{\cG}(\cT) = C_{\cL} \widehat{\cR}_{\cG}(\bm \Pi)$. 

In order to derive $\widehat{\cR}_{\cG}(\bm \Pi)$, we use Dudley's entropy integral \cite[Lemma A.5]{bartlett2017spectrally}, which provides an upper bound on the empirical Rademacher complexity by using the \textit{covering number} of $\bm \Pi$. To clarify, a $\mu$-cover of $\bm \Pi$ is any set $\cC \subseteq \bm \Pi$ 
% set represented by $\cC(\bm \Pi, \mu)$, 
such that $\forall \bpi_{\btheta} \in \bm \Pi$, $\exists \bpi_{\widetilde{\btheta}} \in \cC$ such that 
$$ \max_{\bm \phi \in \cX_{\bm \phi}} \left| \bpi_{\btheta}(\bm \phi) - \bpi_{\widetilde{\btheta}}(\bm \phi) \right| \leq \mu.$$
Similarly, the covering number of the set $\bm \Pi$ at scale $\mu$ is denoted by ${\sf N}(\bm \Pi, \mu)$ and defined as the minimum cardinality of a $\mu$-cover set of $\bm \Pi$. The following lemma summarizes the Dudley's entropy integral that uses the covering number of a set to bound its empirical Rademacher complexity.
\begin{lemma}[{\cite[Lemma~A.5]{bartlett2017spectrally}}]\label{lemma:dudley_integral}
Given samples $\cG$ of size $K$, the empirical Rademacher complexity of the set $\bm \Pi$ with respect to the samples $\cG$ is upperbounded as follows:
\begin{align}\label{eq:dudley_integral}
 \widehat{\cR}_{\cG}(\bm \Pi) \leq \inf_{a > 0} \left( \frac{4 a}{\sqrt{K}} + \frac{12}{K} \int_{a}^{\sqrt{K}} \sqrt{\log{{\sf N}(\bm \Pi, \mu)}} d\mu\right).
\end{align}
\end{lemma}
To proceed with the derivation of $\log({\sf N}(\bm \Pi, \mu))$, we first characterize the Lipschitz constants of the GNN with respect to its parameters. Consider parameters $\btheta$ and $\widetilde{\btheta}$, which correspond to $(\Z_1, \Z_2, \Z_3, \bm \beta)$ and $(\widetilde{\Z}_1, \widetilde{\Z}_2, \widetilde{\Z}_3, \widetilde{\bm \beta})$, respectively. Let $\q_u^{(d)}$ and $\widetilde{\q}_u^{(d)}$ denote the embeddings learned for the $u$th vertex at the end of $d$th layer of the GNN with parameters $\btheta$ and $\widetilde{\btheta}$, respectively. 
% Note that $\q_u^{(0)} = \widetilde{\q}_u^{(0)} = \x_u$ 
Then, for any input $\bm \phi$, the absolute difference between the outputs of the two GNNs can be written as 
\begin{align}\label{eq:lipschitz_output}
    & \left|\bpi_{\btheta}(\bm \phi) - \bpi_{\widetilde{\btheta}}(\bm \phi) \right| \\
    &= \left| \frac{1}{U} \sum_{u \in [U]} \left(\bm \zeta(\bm \beta^\T \q_u^{(D)}) - \bm \zeta(\widetilde{\bm \beta}^\T \widetilde{\q}_u^{(D)}) \right) \right| \nonumber \\ 
    & \leq \frac{1}{U} \sum_{u \in [U]} C_{\bm \zeta} \left|\bm \beta^\T \q_u^{(D)} - \widetilde{\bm \beta}^\T \q_u^{(D)} + \widetilde{\bm \beta}^\T \q_u^{(D)} + \widetilde{\bm \beta}^\T \widetilde{\q}_u^{(D)} \right| \nonumber \\
    & \leq \frac{C_{\bm \zeta}}{U} \sum_{u \in [U]} \left(\left\|\q_u^{(D)}\right\|_2 \left\|\bm \beta - \widetilde{\bm \beta}\right\|_2 + B_{\bm \beta} \left\|\q_u^{(D)} - \widetilde{\q}_u^{(D)} \right\|_2 \right). \nonumber
\end{align}
First, we can bound $\|\q_u^{(D)}\|_2$ as follows:
\begin{align*}
    &\left\|\q_u^{(D)}\right\|_2 \nonumber \\
    &= \Bigg\| \bm \xi \left(\Z_1 \q_u^{(D-1)} + \sum_{(u,v) \in \cE} \bm \xi \left( \Z_2 \q_v^{(D-1)} + \Z_3 \e_{u,v} \right) \right)  - \bm \xi(0) \Bigg\|_2 \nonumber \\
    &\leq C_{\bm \xi} \|\Z_1\|_2 \left\|\q_u^{(D-1)}\right\|_2  \nonumber \\
    & + C_{\bm \xi}^2 \sum_{(u,v) \in \cE}  \bigg( \|\Z_2\|_2 \left\|\q_v^{(D-1)}\right\|_2 + \left\|\Z_3 \right\|_2 \|\e_{u,v}\|_2 \bigg)  \nonumber \\
    &\leq C_{\bm \xi} B_{\Z} \left\|\q_u^{(D-1)}\right\|_2  + C_{\bm \xi}^2 U \max_v \bigg( B_{\Z} \left\|\q_v^{(D-1)}\right\|_2 + B_{\Z} B_{\x} \bigg).
\end{align*}
Solving the recursion from the final inequality, we obtain
\begin{align}
    &\left\|\q_u^{(D)}\right\|_2 \leq \alpha^D B_{\x} + U C_{\bm \xi}^2 B_{\Z} B_{\x} \frac{\alpha^D - 1}{\alpha - 1},
    % &\left\|\q_u^{(D)}\right\|_2 \leq \left((1+UC_{\bm \xi}) C_{\bm \xi} B_{\Z}\right)^D B_{\x} + U C_{\bm \xi}^2 B_{\Z} B_{\x} \frac{\left((1+UC_{\bm \xi}) C_{\bm \xi} B_{\Z}\right)^D - 1}{\left((1+UC_{\bm \xi}) C_{\bm \xi} B_{\Z}\right) - 1},
\end{align}
where $\alpha = ((1+UC_{\bm \xi}) C_{\bm \xi} B_{\Z})$. 

Next, we bound $\Gamma_u^{(D)} := \left\|\q_u^{(D)} - \widetilde{\q}_u^{(D)} \right\|_2$ from \eqref{eq:lipschitz_output} as follows:
\begin{align*}
    & \Gamma_u^{(D)} \nonumber \\
    &= \Bigg\| \bm \xi \left(\Z_1 \q_u^{(D-1)} + \sum_{(u,v) \in \cE} \bm \xi \left( \Z_2 \q_v^{(D-1)} + \Z_3 \e_{u,v} \right) \right) \nonumber\\
    & -  \bm \xi \left(\widetilde{\Z}_1 \widetilde{\q}_u^{(D-1)} + \sum_{(u,v) \in \cE} \bm \xi \left( \widetilde{\Z}_2 \widetilde{\q}_v^{(D-1)} + \widetilde{\Z}_3 \e_{u,v} \right) \right) \Bigg\|_2 \nonumber \\
\end{align*}
\begin{align}
    & \leq C_{\bm \xi} \left\|\Z_1 \q_u^{(D-1)} - \widetilde{\Z}_1 \widetilde{\q}_u^{(D-1)} \right\|_2 \nonumber \\
    & + U C_{\bm \xi}^2 \max_v \left(\left\| \Z_2 \q_v^{(D-1)} - \widetilde{\Z}_2 \widetilde{\q}_v^{(D-1)} \right\|_2 + \left\|\Z_3 - \widetilde{\Z}_3\right\|_2 B_{\x} \right) \nonumber\\
    & \leq C_{\bm \xi} \left( \left\|\q_u^{(D-1)}\right\|_2 \left\|\Z_1 - \widetilde{\Z}_1 \right\|_2 + B_{\Z} \Gamma_u^{(D-1)} \right) \nonumber \\
    & + U \C_{\bm \xi}^2 \max_v \Bigg(\left\|\q_v^{(D-1)}\right\|_2 \left\|\Z_2 - \widetilde{\Z}_2\right\|_2 + B_{\Z} \Gamma_v^{(D-1)} \nonumber \\
    & + B_{\x} \left\|\Z_3 - \widetilde{\Z}_3\right\|_2 \Bigg). \nonumber
\end{align}

Solving the recursion in the last inequality, and using $\Gamma_u^{(0)} = 0, \forall u$, we get
\begin{align*}
    \Gamma_u^{(D)} &\leq \widetilde{\Sigma}_{\Z_1} \left\|\Z_1 - \widetilde{\Z}_1\right\|_2 + \widetilde{\Sigma}_{\Z_2} \left\|\Z_2 - \widetilde{\Z}_2\right\|_2 \nonumber \\
    & + \widetilde{\Sigma}_{\Z_3} \left\|\Z_3 - \widetilde{\Z}_3\right\|_2, \\
    \text{where }\widetilde{\Sigma}_{\Z_1} &=  U C_{\bm \xi}^3 B_{\Z} B_{\x} \frac{\alpha^{(D+1)} - 2 \alpha + 1}{(\alpha - 1)^2}, \\ 
    \widetilde{\Sigma}_{\Z_2} &=  U^2 C_{\bm \xi}^4 B_{\Z} B_{\x} \frac{\alpha^{(D+1)} - 2 \alpha + 1}{(\alpha - 1)^2}, \\
    \widetilde{\Sigma}_{\Z_3} &=  U C_{\bm \xi}^2 B_{\Z} B_{\x} \frac{\alpha^D - 1}{\alpha - 1}.
\end{align*}

Using the above bound on $\Gamma_u^{(D)}$ in \eqref{eq:lipschitz_output}, we get 
\begin{align}\label{eq:final_lipschitz_expression}
     \left| \bpi_{\btheta}(\bm \phi) -  \bpi_{\widetilde{\btheta}}(\bm \phi) \right|  & \leq\Sigma_{\bm \beta} \left\|\bm \beta - \widetilde{\bm \beta}\right\|_2 +\Sigma_{\Z_1} \left\| \Z_1 - \widetilde{\Z}_1\right\|_2 \nonumber\\
    +\Sigma_{\Z_2} & \left\| \Z_2 - \widetilde{\Z}_2\right\|_2  +\Sigma_{\Z_3} \left\| \Z_3 - \widetilde{\Z}_3\right\|_2,
\end{align}
where $\Sigma_{\bm \beta} = C_{\bm \zeta}  B_{\x}\alpha^D + C_{\bm \zeta} U C_{\bm \xi}^2 B_{\Z} B_{\x} \frac{\alpha^D - 1}{\alpha - 1}$, $\Sigma_{\Z_1} = C_{\bm \zeta} B_{\bm \beta} \widetilde{\Sigma}_{\Z_1}$, $\Sigma_{\Z_2} = C_{\bm \zeta} B_{\bm \beta} \widetilde{\Sigma}_{\Z_2}$, and $\Sigma_{\Z_3} = C_{\bm \zeta} B_{\bm \beta} \widetilde{\Sigma}_{\Z_3}$.

Eq.~\eqref{eq:final_lipschitz_expression} implies that for any $\btheta \in \bm \varTheta$, the existence of $\widetilde{\btheta}$ in the cover set such that $|\bpi_{\btheta}(\bm \phi) -  \bpi_{\widetilde{\btheta}}(\bm \phi)| \leq \mu$ can be satisfied by ensuring the existence of $(\widetilde{\bm \beta}, \widetilde{\Z}_1, \widetilde{\Z}_2, \widetilde{\Z}_3)$ such that the right hand side of \eqref{eq:final_lipschitz_expression} $\leq \mu$. Hence, if we construct $\nicefrac{\mu}{4\Sigma_{\bm \beta}}$-cover of $\cX_{\bm \beta}$, and $\nicefrac{\mu}{4\Sigma_{\Z_i}}$-cover of $\cX_{\Z}$, $\forall i \in \{1,2,3\}$, the Cartesian product of the four sets correspond to a $\mu$-cover of $\bm \Pi$. Hence, the covering number of $\bm \Pi$ at scale $\mu$ can be upper bounded by the product of the covering numbers of the four sets as follows:
\begin{align}\label{eq:combined_covering_number}
 {\sf N}\left(\bm \Pi, \mu \right) \leq {\sf N}\left(\cX_{\bm \beta}, \frac{\mu}{4\Sigma_{\bm \beta}}\right) \times \prod_{i=1}^3 {\sf N} \left(\cX_{\Z}, \frac{\mu}{4\Sigma_{\Z_i}}\right).
\end{align}
In addition, the covering number for $\cX_{\Z}$ and $\cX_{\bm \beta}$  can be upper bounded using \cite[Lemma 8]{chen2019generalization} and \cite{pollard1990empirical}, respectively, as follows:
$$ {\sf N}(\cX_{\Z}, \mu) \leq \left( 1 + \frac{2\sqrt{E} B_{\Z}}{\mu} \right)^{E^2}, {\sf N}(\cX_{\bm \beta}, \mu) \leq \left( \frac{3 B_{\bm \beta}}{\mu}\right)^E$$
% Similarly, the covering number for $\cX_{\bm \beta}$ can be upper bounded as follows \cite{pollard1990empirical}:
% $$ {\sf N}(\cX_{\bm \beta}, \mu) \leq \left( \frac{3 B_{\bm \beta}}{\mu}\right)^E.$$
Using the above bounds in \eqref{eq:combined_covering_number}, we get 
\begin{align*}
    & {\sf N} \left(\bm \Pi, \mu \right) \leq  \left( \frac{12 B_{\bm \beta}\Sigma_{\bm \beta}}{\mu}\right)^E \times \prod_{i=1}^3 \left( 1 + \frac{8\sqrt{E} B_{\Z}\Sigma_{\Z_i}}{\mu} \right)^{E^2} \\
    &  \leq \left(1 + \frac{12 \sqrt{E} B_{\Z} {\rm max}\left\{ \frac{B_{\bm \beta}}{B_{\Z}}\Sigma_{\bm \beta},\Sigma_{\Z_1},\Sigma_{\Z_2},\Sigma_{\Z_3}\right\}}{\mu} \right)^{3E^2 + E}. 
\end{align*}

Finally, we can use Lemma \ref{lemma:dudley_integral} to obtain a bound on $\widehat{\cR}_{\cG}(\bm \Pi)$. 
% \begin{align*}
%     \widehat{\cR}_{\cG}(\bm \Pi) & \leq \inf_{a > 0} \left( \frac{4 a}{\sqrt{K}} + \frac{12}{K} \int_{a}^{\sqrt{K}} \sqrt{\log{{\sf N}(\bm \Pi, \mu)}} d\mu\right) 
% \end{align}
To this end, we upper bound the integral on the right hand side of \eqref{eq:dudley_integral} as follows:
\begin{align*}
    \int_{a}^{\sqrt{K}} \sqrt{\log{{\sf N}(\bm \Pi, \mu)}} d\mu ~\leq~ \sqrt{K} \sqrt{\log{{\sf N}(\bm \Pi, a)}}.
\end{align*}
The above inequality holds because $\sqrt{\log{{\sf N}(\bm \Pi, \mu)}}$ increases monotonically with the decrease of $\mu$. %Hence the maximum value within the range $a$ and $\sqrt{K}$ is $\sqrt{\log{{\sf N}(\bm \Pi, a)}}$, whereas the maximum range of integration is $[0, \sqrt{K}]$. 
Taking $a= \nicefrac{1}{\sqrt{K}}$, we get the following:
\begin{align}
  &\widehat{\cR}_{\cG}(\bm \Pi) \leq  \frac{4}{K} + \frac{12 \sqrt{3E^2 + E}}{\sqrt{K}} \times \nonumber \\
  &  \sqrt{ \log\left(1 + {12 \sqrt{EK} B_{\Z} {\rm max}\left\{ \frac{B_{\bm \beta}}{B_{\Z}}\Sigma_{\bm \beta},\Sigma_{\Z_1},\Sigma_{\Z_2},\Sigma_{\Z_3} \right \}}\right)}. \nonumber
\end{align}
Combining the above with $\widehat{\cR}_{\cG}(\cT) \leq C_{\cL}\widehat{\cR}_{\cG}(\bm \Pi)$ and substituting in \eqref{eq:generalization_error}, we get the final result.

		\section{Proof of Theorem \ref{thm:main}}\label{app:proof_ml}

Proof of Theorem \ref{thm:main} can be divided into two parts. In the first part we bound the expected loss under of the learned GNN. For this we will use the proof idea from \cite{ross2011reduction}. However, the proof technique in \cite{ross2011reduction} hinges on the convexity of their online learning problem.
Hence, we make appropriate modifications to accommodate our non-convex GNN-based learning problem. In the second part, using the expected loss, we characterize the number of nodes needed to be visited by Algorithm \ref{algo:abb} for solving a given problem instance optimally. 

\subsection{Expected Loss of Algorithm \ref{algo:prune_learning}}
Note that the online learning algorithm in Algorithm~\ref{algo:prune_learning} is a no-regret algorithm.
The definition of regret is as follows:
\begin{definition}[Regret]
 Regret of an online algorithm that produces a sequence of policies  $\btheta_{1:I} = \{\btheta^{(1)}, \btheta^{(2)}, \dots, \btheta^{(I)}\}$ is denoted by ${\rm Reg}_I$. It is the average loss of all policies with respect to the best policy in hindsight, i.e.,
\begin{align*}
 {\rm Reg}_I &:= \frac{1}{I}\sum_{i=1}^I \frac{1}{|\cD_i|} \sum_{(\bm \Phi_s, y_s) \in \cD_i} [ \cL(\bpi_{\btheta^{(i)}}(\bm \phi_s), y_s)]  \\
 & - \min_{\btheta \in \bm \varTheta} \frac{1}{I}\sum_{i=1}^I \frac{1}{|\cD_i|} \sum_{(\bm \Phi_s, y_s) \in \cD_i} [ \cL(\bpi_{\btheta}(\bm \phi_s), y_s)].
\end{align*}
\end{definition}
\begin{definition}[No-regret Algorithm]
A no-regret algorithm is an algorithm that produces a sequence of policies $\btheta_{1:I}$ such that the average regret goes to $0$ as $N$ goes to $\infty$:
$$ {\rm Reg}_I \leq \gamma_I \quad \text{and} \quad \lim_{I \to \infty} \gamma_I \to 0 .$$
\end{definition}
% With the introduction of the random regularization $\bm r(\bm \theta)=-\bm \psi^\T\bm \theta$, Algorithm \ref{algo:prune_learning} is an instance of an online learning algorithm called \textit{Follow The Perturbed Leader} (FTPL) \reminder{cite}. 
For strongly convex $\cL$, the work in \cite{ross2011reduction} shows that Algorithm \ref{algo:prune_learning} is a no-regret algorithm with $\eta = \infty$, i.e., $\bm \psi = \zero$ (recall that $\eta$ is the parameter of the exponential distribution, i.e., $\bm \psi \sim {\rm Exp}(\eta)$, where ${\rm Exp}(\eta) := \eta(\exp(-\eta))$). However, for non-convex $\cL$ we cannot guarantee that Algorithm \ref{algo:prune_learning} is a no-regret algorithm \cite{agarwal2019learning}. But with $ 0 < \eta < \infty$, under Assumption \ref{ass:lipschitz}, Algorithm \ref{algo:prune_learning} was shown to be a no-regret algorithm \cite{agarwal2019learning}.  
\begin{lemma}\label{lemma:regret} {\cite[Theorem 1]{agarwal2019learning}}
When Assumption \ref{ass:lipschitz} holds, the regret after $N$ iterations can be bounded by:
$$ \bbE_{\bm \psi \sim {\rm Exp}(\eta)} [{\rm Reg}_I] \leq \gamma_I \leq \cO(1/I^{1/3}).$$
\end{lemma}
Finally, the following lemma establishes the expected loss of the policy returned by Algorithm \ref{algo:prune_learning}.
\begin{lemma}\label{lemma:error_bound}
For Algorithm \ref{algo:prune_learning}, with probability at least $1-\delta$,
\begin{align}
& \min_{\btheta \in \btheta_{1:I}} \bbE_{(\bm \phi_s, y_s) \sim p_{\btheta}, \bm \psi}[ \cL(\bpi_{\btheta}(\bm \phi_s), y_s)] \nonumber \\
& \leq  \min_{\btheta \in \bm \varTheta} \frac{1}{I} \sum_{i=1}^I \frac{1}{J} \sum_{(\bm \phi_s,y_s) \in \cD_i} \bbE_{\bm \psi}[\cL(\bpi_{\btheta}(\bm \phi_s), y_s)] \nonumber \\
& + \gamma_I + {\sf Gap} \left(\frac{\delta}{2}, J\right) \sqrt{\frac{2\log(\frac{2}{\delta})}{I}}.
\end{align}
\end{lemma}

\begin{IEEEproof}
Define $\omega_{i}, \forall i \in [I]$ as:
\begin{align*}
 \omega_{i} := & \bbE_{p_{\btheta^{(i)}}, \bm \psi}[\cL(\bpi_{\btheta^{(i)}}(\bm \phi_s), y_s)] \\
 & - \frac{1}{J} \sum_{(\bm \phi_s, y_s) \in \cD_{i}} \bbE_{\bm \psi}[ \cL(\bpi_{\btheta^{(i)}}(\bm \phi_s), y_s)]. 
\end{align*}
Next, we use Lemma~\ref{lemma:generalizatio_error} to obtain a bound on $\omega_i, \forall i$; i.e., with probability at least $1-\nicefrac{\delta}{2}$, the following holds simultaneously for $\omega_i, \forall i \in [I] $ :
$ \omega_i \leq {\sf Gap}\left(\frac{\delta}{2}, J\right).$
Consequently, $\Omega_i := \sum_{t=1}^i \omega_{t}, i=\{1, \dots, I\}$ forms a martingale sequence, i.e., 
$\bbE[\Omega_i | \Omega_1, \dots, \Omega_{i-1}] = \Omega_{i-1}.$
Also, we have $|\Omega_{i+1} - \Omega_{i}| \leq {\sf Gap}(\nicefrac{\delta}{2}, J), \forall i \in [I-1]$ with probability $1-\nicefrac{\delta}{2}$. Next, consider the following lemma:
\begin{lemma}[Azuma-Hoeffding's Inequality]\label{lemma:azuma}
    Let $X_0, \dots, X_I$ be a martingale sequence and $|X_{i} - X_{i-1}| \leq c_i $. Then with probability $1-\delta$, 
    \begin{align*}
        {\rm Pr}(X_I - X_0 \geq \epsilon) \leq \exp\left( \frac{- \epsilon^2}{2 \sum_{i=1}^I c_i^2} \right).
    \end{align*}
\end{lemma}
Using Lemma \ref{lemma:azuma}, we have the following holds with probability of at least $(1-\nicefrac{\delta}{2})^2 \geq 1- \delta$,
\begin{align}\label{eq:omega_bound}
\Omega_{I} \leq {\sf Gap} \left(\frac{\delta}{2}, J\right)\sqrt{2 I \log(2/\delta)}.
\end{align}

% Let $\cD_{ij}$ denote the empirical distribution of states visited by $\bpi_{\btheta^{(i)}}$ in trajectory $j$. Define $Y_{ij}$ as:
% $$ Y_{ij} =\bbE_{p_{\btheta^{(i)}}, \bm \psi}[\cL(\bpi_{\btheta^{(i)}}(\bm \phi_s), y_s)] - \frac{1}{|\cD_{ij}|} \sum_{(\bm \phi_s, y_s) \in \cD_{ij}} \bbE_{\bm \psi}[ \cL(\bpi_{\btheta^{(i)}}(\bm \phi_s), y_s)]. $$
% We know that $Y_{ij}$ is a zero mean random variable bounded within $[-1, 1 ]$. {\orange This assumption seems problematic for binary cross entropy loss}

% The ordered sequence $Y_{11}, Y_{12},\dots, Y_{1J}, Y_{21}, \dots, Y_{IJ}$ forms a martingale sequence. 
% By Azuma-Hoeffding's inequality, with probability $1-\delta$,
% $$ \frac{1}{IJ} \sum_{i=1}^I \sum_{j=1}^J Y_{IJ} \leq \sqrt{\frac{2 \log(1/\delta)}{I}}.$$
% {\orange [This is a bound on the empirical and expected loss which can be replaced by rademacher complexity based generalization error bound]}

% Therefore, with probability $1-\delta$,
Now, consider the following inequality:
\begin{align*}
    & \min_{\btheta \in \btheta_{1:I}} \bbE_{p_{\btheta}, \bm \psi}[ \cL(\bpi_{\btheta}(\bm \phi_s), y_s)] \\
    & \leq \frac{1}{I} \sum_{i=1}^I \bbE_{p_{\btheta_{i}}}\bbE_{\bm \psi}[\cL(\bpi_{\btheta^{(i)}}(\bm \phi_s), y_s)] \\
    &= \frac{1}{I} \sum_{i=1}^I  \frac{1}{J} \sum_{(\bm \phi_s, y_s) \in \cD_i}\bbE_{\bm \psi} [ \cL(\bpi_{\btheta^{(i)}}(\bm \phi_s), y_s)] +  \frac{1}{I} \sum_{i=1}^I \omega_{i}.
\end{align*}
Hence, with probability of at least $1-\delta$, we have
\begin{align*}
    & \min_{\btheta \in \btheta_{1:I}} \bbE_{p_{\btheta}, \bm \psi}[ \cL(\bpi_{\btheta}(\bm \phi_s), y_s)] \\
    & \stackrel{(a)}{\leq} \min_{\btheta \in \bm \varTheta} \frac{1}{I}\sum_{i=1}^I \frac{1}{J} \sum_{(\bm \phi_s, y_s) \in \cD_i}\bbE_{\bm \psi} [ \cL(\bpi_{\btheta}(\bm \phi_s), y_s)] + \cO(1/I^{1/3}) \\
    & + {\sf Gap} \left(\frac{\delta}{2}, J\right)\sqrt{\frac{2 \log(2/\delta)}{I}} \\
\end{align*}
\begin{align*}
    & \stackrel{(b)}{\leq} \nu + \cO(1/I^{1/3}) + {\sf Gap} \left(\frac{\delta}{2}, J\right) \sqrt{\frac{2 \log(2/\delta)}{I}},
\end{align*}
where (a) is by Lemma~\ref{lemma:regret} and \eqref{eq:omega_bound}, and (b) is obtained via using Assumption \ref{ass:realizability}.
\end{IEEEproof}

When the loss function $\cL$ is selected to be binary cross-entropy loss, i.e., 
$$ \cL(x,y) = -y\log(x) - (1-y)\log(1-x),$$
$1-e^{-\cL(x,y)}$ corresponds to the classification error. Therefore, classification accuracy for any $\btheta$, i.e., $\rho_{\btheta}$ is given by 
$$ \rho_{\btheta} = \bbE_{p_{\btheta}, \bm \psi} [ \exp(-\cL(\bpi_{\btheta}(\bm \phi_s), y_s)) ]. $$
%{\red a little lost here. should we use the empirical version? as $\rho_{\bm \theta}$? Maybe not, since in the end we characterize the expected number of nodes. Just a comment for myself.} {\orange $\exp(-L(\bpi_{\btheta}(\bm \phi),y))$ gives the probability of not making a mistake for $\bm \phi,y$. So, we need to take expectation over the distribution of $\bm \phi, y$. $\bm \psi$ does not actually affect anything here. }

Note that $\widehat{\btheta} = \arg \min_{\btheta \in \btheta_{1:I}} \bbE_{p_{\btheta}, \bm \psi}[ \cL(\bpi_{\btheta}(\bm \phi_s), y_s)]$. 
Next, we characterize $\rho_{\widehat{\btheta}}$. 
To that end, the following follows from Lemma~\ref{lemma:error_bound}.
\begin{align*}
    & \exp(\bbE_{p_{\btheta}, \bm \psi}[-\cL(\bpi_{\btheta}(\bm \phi_s), y_s)]) \\
    & \geq \exp\left( - \nu - \cO(1/I^{1/3}) - {\sf Gap} \left(\frac{\delta}{2}, J\right)\sqrt{\frac{2 \log(2/\delta)}{I}} \right) \\ 
    \implies & \rho_{\widehat{\btheta}} = \bbE_{p_{\btheta}, \bm \psi} [ \exp(-\cL(\bpi_{\btheta}(\bm \phi_s), y_s)) ] \\
    & \stackrel{(b)}{\geq} \exp\left( - \nu - \cO(1/I^{1/3}) - {\sf Gap} \left(\frac{\delta}{2}, J\right)\sqrt{\frac{2 \log(2/\delta)}{I}} \right),
\end{align*}
where (b) follows from Jensen's inequality.

\subsection{B\&B expected number of nodes and optimality}
{Let $\epsilon_{\rm FP}$ denote the false positive error rate, i.e.,  the probability of classifying an irrelevant node as relevant. Also define $\epsilon_{\rm FN}$ denote the false negative error rate, i.e., the probability of classifying a relevant node as irrelevant.} Then the expected number of branches generated by using pruning policy on B\&B was derived in \cite{he2014learning}:

\begin{lemma}[{\cite[Theorem 1]{he2014learning}}]\label{lemma:expected_branches}
Assume that the node selection method in \eqref{eq:node_selection} ranks an irrelevant node higher than a relevant node with probability $\epsilon_r$. Then the expected number of branches (number of non-leaf nodes) is 
\begin{align*}
 \frac{Q_{\widehat{\btheta}} - 1}{2}
 \leq \bigg( \left( \frac{1-\epsilon_{\rm FN}}{1- 2 \epsilon_r \epsilon_{\rm FP}} + \frac{\epsilon_{\rm FN}}{1-2\epsilon_{\rm FP}} \right)\epsilon_r \epsilon_{\rm FP} \sum_{n=0}^N (1-\epsilon_{\rm FN})^n  \\
+ (1-\epsilon_{\rm FN})^{N+1} \frac{(1-\epsilon_r)\epsilon_{\rm FP}}{1-2\epsilon_{\rm FP}} + 1 \bigg)N,
\end{align*}
\end{lemma}
% {\red again, what is the expectation taken over?} {\orage described in the main theorem}
% We have assumed that $\rho_{\widehat{\btheta}} = 1- \epsilon_{\rm FP} = 1 - \epsilon_{\rm FN}$. {\red we have assumed, or in our case, we have the equality?} {\orange Here we are making a simplifying assumption because of two reasons, (a) we can not obtain $\epsilon_{\rm FN}$ and $\epsilon_{\rm FP}$ by using $\cL(\bpi_{\btheta}(\bm \phi), y)$ (b) to get a simple expression for $Q_{\widehat{\btheta}}$ that is easy to interpret.
% }

Our node selection strategy is the lowest lower bound first as detailed in Section \ref{sec:bb}. In the worst case scenario, $\epsilon_r = 1$. Therefore, using Lemma \ref{lemma:expected_branches}, the expected number of branches is
\begin{align*}
 & \leq N \left( \frac{1 - \rho_{\widehat{\btheta}}}{2 \rho_{\widehat{\btheta}} - 1} \sum_{n=0}^N \rho_{\widehat{\btheta}}^n  + 1 \right) \stackrel{(c)}{=} N \left( \frac{1 - \rho_{\widehat{\btheta}}^{N+1}}{2 \rho_{\widehat{\btheta}} - 1} + 1\right) \\
 & = \frac{ N (2 \rho_{\widehat{\btheta}} - \rho_{\widehat{\btheta}}^N)}{2 \rho_{\widehat{\btheta}} - 1}.
\end{align*}
Since the expected number of branches correspond to the expected number of non-leaf nodes, the total number of nodes in the tree is 
$ \leq \frac{2N (2 \rho_{\widehat{\btheta}} - \rho_{\widehat{\btheta}}^N)}{2 \rho_{\widehat{\btheta}} - 1} + 1.$
Next, we we characterize the probability that Algorithm \ref{algo:abb} provides the optimal solution. To this end, observe that there is only one relevant node at any depth $n$ of the B\&B algorithm. The probability of not pruning a relevant node is $\geq \rho_{\widehat{\btheta}}$. Therefore, the probability of not pruning a relevant node at any depth of the branch and bound tree is $\geq \rho_{\widehat{\btheta}}^N$ (since $N$ is the maximum depth of the tree). Hence, the probability of obtaining an optimal solution is at least $\rho_{\widehat{\btheta}}^N$.

		\section{GNN Design in Simulations} \label{app:gnn}
		In this section, we detail the GNN architecture used in the experiments. The GNN is designed to accommodate the unequal input feature dimensions for antennas and users. 
		We enhance the expressiveness GNN by letting different layers to have different aggregation matrices in our experiments.
		The initial embeddings of a common size $E$ are obtained using a single layer fully connected neural network, i.e., 
		\begin{align*}
			\q^{(0)}_n &= {\sf ReLU}(\Z_1 \x_n), ~ \q^{(0)}_{N+m} = {\sf ReLU}(\Z_2 \x_m) \\
			\e_{u,v} &= {\sf ReLU}(\Z_3 \widetilde{\e}_{u,v}).
		\end{align*}
		where $\Z_1 \in \bbR^{E \times V_a}$, $\Z_2 \in \bbR^{E \times V_u}$, $\Z_3 \in \bbR^{E \times V_e}$, and ${\sf ReLU}: \bbR^E \to \bbR^E$ deontes elementwise nonlinear function such that ${\sf ReLU}(x) = {\rm max}\{x,0\}$. 
		
		The first layer of GNN only updates the antenna vertices, i.e., $\q_{n}, n \in [N]$, as follows
		\begin{align*}
			\q_{n}^{(1)} &= \Z_9\Bigg({\sf ReLU} \Bigg( \Z_8 \q_n^{(0)} + \sum_{m=1}^M \Z_7 \Big({\sf ReLU} \Big(\Z_6 \q_n^{(0)} + \\
			& \Z_5 \q_{N+m}^{(0)} + \Z_4 \e_{n, N+m}\Big) \Big) \Bigg) \Bigg), \forall n \in [N] \nonumber \\
			\q_{N+m}^{(1)} &= \q_{N+m}^{(0)}, \forall m \in [M].
		\end{align*}
		The second layer only updates the user vertices as follows
		\begin{align*}
			\q_{N+m}^{(2)} &= \Z_{15}\Bigg({\sf ReLU} \Bigg( \Z_{14} \q_{N+m}^{(1)} + \sum_{n=1}^N \Z_{13} \Big({\sf ReLU} \Big(\Z_{12} \q_{N+m}^{(1)}\\
			&+ \Z_{11} \q_{n}^{(1)} + \Z_{10} \e_{n, N+m}\Big) \Big) \Big) \Big), \forall m \in [M] \nonumber \\
			\q_{n}^{(2)} &= \q_{n}^{(1)}, \forall n \in [N].
		\end{align*}
		Such ``split updating'' of different nodes' embeddings in two layers has been advocated in \cite{gasse2019exact} for the type of graph structure used in this work (i.e., a bipartite graph). Moreover, there is a potential saving in the computational cost in both training and testing  \cite{nassar2018hierarchical} compared to updating all nodes' embeddings in each layer.
		
		Finally, $\bpi_{\btheta}(\bm \phi)$ is computed using the $\q_{N+m}^{(2)}, \forall m \in [M]$ as follows:
		$$ \bpi_{\btheta}(\bm \phi) = {\sf Sigmoid} \left( \frac{1}{M} \sum_{m=1}^M \bm \beta^\T {\sf ReLU}(\Z_{16} \q_{N+m}^{(2)}) \right),$$
		where $\Z_4, \dots \Z_{16} \in \bbR^{E \times E}$, $\bm \beta \in \bbR^E$, and ${\sf Sigmoid}: \bbR \to \bbR$ is the sigmoid function, i.e., ${\sf Sigmoid}(x) = \frac{1}{1 + \exp(-x)}$.

		\section{Construction of input features ($\bm \phi ( \cN_s^{(\ell)})$)}\label{sec:app_feature_implementation}
		
		We assign the features tabulated in Table~\ref{tab:feature_design} among the elements of the following sets: $\{\x_i ~|~ i \in [N]\}$, $\{\x_{N+i} ~|~ i \in [M]\}$, and $\{\e_{i,N+j} ~|~ i \in [N], j \in [M]\}$. Specifically, the {Type II} features that can be represented with a vector of dimension $N$ (i.e., $\cA_s^{(\ell)}$, and $\cB_s^{(\ell)}$, $[\|\W_{\ell, s}(1,:)\|_2^2, \dots, \|\W_{\ell, s}(N,:)\|_2^2]$) are assigned to the elements of $\{\x_i ~|~ i \in [N]\}$ as follows:
		\begin{align*}
			x_i(1) &= \begin{cases} 1, \text{ if } i \in \cA_s^{(\ell)} \\ 0, \text{ otherwise,} \end{cases} ~ x_i(2) = \begin{cases} 1, \text{ if } i \in \cB_s^{(\ell)} \\ 0, \text{ otherwise, and} \end{cases} \\
			x_i(3) &= \|\W_{\ell, s}(i,:)\|_2^2.
		\end{align*}
		Similarly, the {Type II} features that can be represented by a vector of dimension $M$ (i.e., $\W_{\ell,s}(:,m)^H \h_m$ and the aggregated interference under $\W_{\ell,s}$) are assigned to be the elements of $\{\x_{N+i} ~|~ i \in [M]\}$ as follows:
		\begin{align*}
			x_{N+i}(1) = \left|\W_{\ell,s}(:,i)^H \h_i\right|^2, ~x_{N+i}(2) = \sum_{j\neq i} \left|\W_{\ell,s}(:,j)^H \h_i\right|^2.
		\end{align*}
		The remaining {Type II} features can be represented by a vector of dimension $NM$, and are assigned to the elements of $\{\e_{i,N+j} ~|~ i \in [N], j \in [M]\}$ as follows:
		\begin{align*}
			&(e_{i,N+j}(1), e_{i,N+j}(2), e_{i,N+j}(3)) = ({\rm Re}(\H(i,j)), \\
			& \quad \quad \quad \quad \quad  {\rm Im}(\H(i,j)), |\H(i,j)|) \\
			& (e_{i,N+j}(4), e_{i,N+j}(5), e_{i,N+j}(6)) = ({\rm Re}(\W_{\rm incumbent}(i,j)), \\
			& \quad \quad \quad \quad \quad {\rm Im}(\W_{\rm incumbent}(i,j)), |\W_{\rm incumbent}(i,j)|) \\
			& (e_{i,N+j}(7), e_{i,N+j}(8), e_{i,N+j}(9)) = ({\rm Re}(\W_{\ell,s}(i,j)),\\
			& \quad \quad \quad \quad \quad  {\rm Im}(\W_{\ell,s}(i,j)), |\W_{\ell,s}(i,j)|), 
		\end{align*}
		where ${\rm Re}(\cdot)$ and ${\rm Im}(\cdot)$ returns the real and imaginary part of the complex number.
		
		Finally, the {Type I} features are assigned to the set $\{ \x_{N+i} ~|~ i \in [M]\}$ as follows:
		\begin{align*}
			& (x_{N+i}(3), x_{N+i}(4), \dots, x_{N+i}(8)) \\
			& = (l_G^{(t)}, u_G^{(t)}, \Phi_{\rm lb}(\cN_s^{(\ell)}), \Phi_{\rm ub}(\cN_s^{(\ell)}), \ell, \mathbb{1}(\Phi_{\rm ub}(\cN_s^{(\ell)}) - u_G^{(t)} < \epsilon)).
		\end{align*}
		

		\ifCLASSOPTIONcaptionsoff
		\newpage
		\fi
	\end{document}